\documentclass[12pt,preprint]{aastex}

\newcommand{\feoh}{[{\rm Fe} / {\rm H}]}
\newcommand{\msun}{\, M_\odot}

\def\nuc#1{${}^{#1}$}
\def\nucm#1#2{{}^{#1}{\rm #2}}
\def\abra#1#2{{\rm #1}/ {\rm #2}}
\shorttitle{Carbon-Enhancement and IMF of EMP-Stars}
\shortauthors{Komiya et al.}

\begin{document}

\title{The Origin of Carbon-Enhancement and \\ Initial Mass Function of \\ Extremely Metal-Poor Stars in the Galactic Halo}

\author{Yutaka Komiya\altaffilmark{1}, Takuma Suda \altaffilmark{1,2}, Hiroki Minaguchi \altaffilmark{3,4}, \\ Toshikazu Shigeyama \altaffilmark{4}, Wako Aoki \altaffilmark{5}, and Masayuki Y. Fujimoto \altaffilmark{1} }
\altaffiltext{1}{Department of Physics, Faculty of Science, Hokkaido University, Sapporo, Hokkaido 060-0810, Japan}
\altaffiltext{2}{Meme Media Laboratory, Hokkaido University, Sapporo, Hokkaido 060-0813, Japan }
\altaffiltext{3}{Department of Astronomy, Graduate School of Science, University of Tokyo, Bunkyo-ku, Tokyo 113-0033, Japan}
\altaffiltext{4}{Research Center for the Early Universe, Graduate School of Science, University of Tokyo, Bunkyo-ku, Tokyo 113-0033, Japan}
\altaffiltext{5}{National Astronomical Observatory, Mitaka, Tokyo, 181-8588 Japan}

\begin{abstract} 
It is known that the carbon-enhanced, extremely metal-poor (CEMP) stars constitute a substantial proportion in the extremely metal-poor (EMP) stars of the Galactic Halo, by far larger than CH stars in Population II stars.  
   We investigate their origin with taking into account an additional evolutionary path to the surface carbon-enrichment, triggered by hydrogen engulfment by the helium flash convection, in EMP stars of $\feoh \lesssim -2.5$.  
  This process is distinct from the third dredge-up operating in more metal-rich stars and also in EMP stars.  
   In binary systems of EMP stars, the secondary stars become CEMP stars through mass transfer from the primary stars of low and intermediate masses, which have developed the surface carbon-enhancement.  
   Our binary scenario can predict the variations in the abundances not only for carbon but also for nitrogen and s-process elements and reasonably explain the observed properties such as the stellar distributions with respect to the carbon abundances, the binary periods, and the evolutionary stages.  
   Furthermore, from the observed frequencies of CEMP stars with and without s-process element enhancement, we demonstrate that the initial mass function of EMP stars need to give the mean mass $\sim 10\msun$ under the reasonable assumptions on the distributions of orbital separations and mass ratio of binary components. 
   This also indicates that the currently observed EMP stars were exclusively born as the secondary members of binaries, making up $\sim 10\%$ remnants of EMP binary systems of mass $\sim 10^8 \msun$ in total;  
   in addition to CEMP stars with white dwarf companions, a significant fraction of them have experienced supernova explosions of their companions.  
   We discuss the implications of the present results in relation to the formation of Galactic halo. 
\end{abstract}
\keywords{stars: abundances --- stars: carbon --- Galaxy: halo --- stars: evolution --- stars: luminosity function, mass function}

\section{Introduction}\label{introduction}

\begin{figure}
\epsscale{1.0} \includegraphics{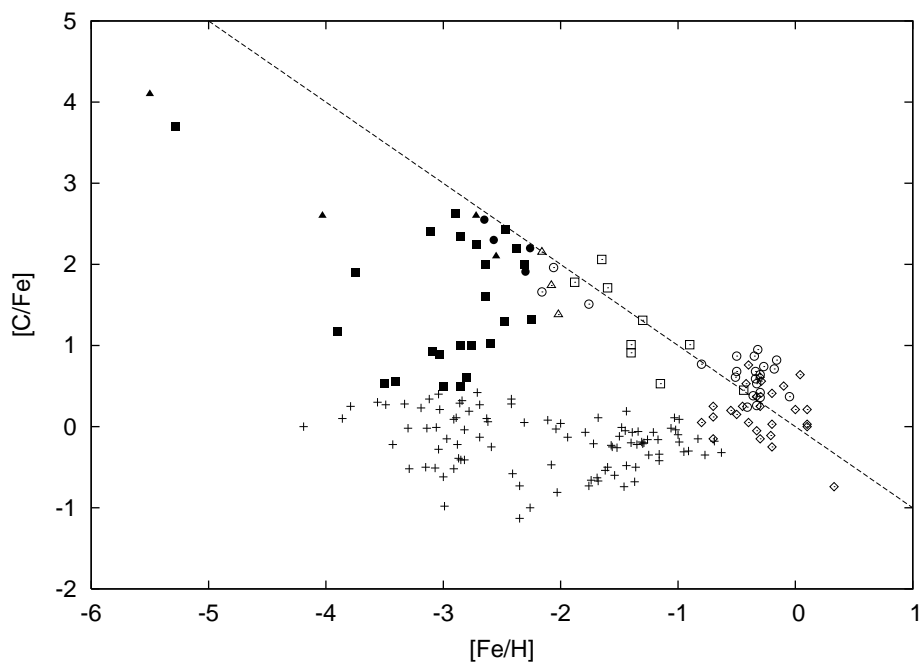}
\caption{Sample stars compiled from literature of abundance analysis with high dispersion spectroscopy. 
   Filled symbols denote carbon-enhanced, extremely metal-poor (CEMP) stars, divided into three evolutionary stages of giants (squares), subgiants (circles), and dwarfs (triangles).
   Open symbols denote CH stars, divided into four groups of CH giants (squares), Ba stars (diamonds), subgiant CH stars including both dwarfs and subgiants (circles) and field blue stragglers showing strong CH lines (triangles). 
   Broken line denotes the locus of $[\abra{C}{H}] = 0$.  
   Crosses denote stars without carbon-enhancement, though our compilation is not complete for stars of $\feoh > -2.5$ and for $[\abra{C}{Fe}] < 0.5$. }
\label{FeC}
\end{figure}

Both HK survey \citep{Beers92} and Hamburg/ESO (HES) survey \citep{Christlieb01} have shown that a large fraction ($20 \sim 25$ \%) of extremely metal-poor (EMP) stars ($\feoh \lesssim -2.5$) exhibit the enhancement of surface carbon abundance. 
\citep[][see also Beers \& Christlieb 2005]{Rossi99, Beers99,Norris01, Christlieb03}.  
   This forms a striking contrast to the fact that more metal-rich CH stars, characterized by enhanced CH bands in the spectra, account for only $\sim 1 \%$ of the other Population II (Pop II) stars \citep{Tomkin89,Luck91}.
   Figure~\ref{FeC} shows the observed carbon enrichment against the metallicity for the stars whose abundances are available from the analyses of high dispersion spectroscopy.  
   This figure contains 34 CEMP stars for $\feoh \lesssim -2.5$, and we can readily see the prominence of CEMP stars among EMP stars though the sample may not be free from bias. 
   In addition, the various peculiar aspects have been revealed; 
   a) CEMP stars are divided into two groups, one (CEMP-s) with the enhancement of s-process element up to $2 \sim 3$ dex and the other (CEMP-nos) with normal s-process element abundances \citep[][see also Figure~\ref{BaFeNCFe} below]{Norris97a,Aoki02a,Ryan05}: 
  b) a fairly large group of stars display great enrichment of nitrogen relative to carbon \citep{Norris02,Spite05}:
  and c) a bunch of stars show the significant enhancement of r-process elements \citep{Sneden96} with large variations in the degree up to a factor of $\sim 100$ or more \citep{Honda04}.  

These features may arise from the peculiarities in the structure and evolution of stars due to very low metallicity and also from their formation processes, characteristic to the early stage of our Galaxy. 
   Theoretically, it has been known that low-mass stars of metallicity $\feoh \lesssim - 2.5$ dredge up carbon to the surface by a mechanism different from that in the stars of the metal-rich populations \citep{Fujimoto90,Fujimoto00}. 
   This mechanism, which is called helium-flash driven deep mixing (He-FDDM; see \S 2), is proposed to explain the observed features of carbon enhanced EMP (CEMP) stars.  
   Because of this theoretical property as well as of the observational characteristics, above mentioned, we call stars of $\feoh \lesssim -2.5$ as EMP stars in the following, and distinguish them from ``Pop II" stars of $ \feoh \gtrsim -2.5 $. 

There is a well-established scenario for the origin of carbon-rich stars among the Pops I and II stars. 
   According to \citet{Iben75} and \citet{Iben83b}, a Pop I or II star enhances the carbon abundance in the hydrogen-rich envelope by the third dredge-up (TDU) when it evolves to the asymptotic giant branch (AGB). 
   Then the star loses its envelope by the stellar wind, eventually turning into a white dwarf. 
   If this star belongs to a binary system, the companion star, less massive than this AGB star initially, is exposed to, and accretes, the wind matter enriched with carbon and changes its surface composition to exhibit the carbon enrichment. 
   As a result, a carbon star remains with an unseen companion star (a white dwarf). 
   This is supported by the results of spectroscopic observations that all CH stars belong to binary systems with unseen companions \citep{McClure90}.  

We propose that all CEMP stars are also made in the binaries similarly to CH stars, but with carbon dredged up by He-FDDM as well as by TDU. 
   A statistical analysis of carbon-enhanced EMP stars enriched with the s-process elements (CEMP-s) by \citet{Lucatello05a} suggests that most of these stars belong to multiple systems, supporting the argument that CEMP-s stars are metal-poor analogs of CH stars. 
   In fact, if plotted as a function of the surface carbon enhancement ($[\abra{C}{H}]$), CEMP dwarfs and CEMP giants have distinct distributions, as shown in Figure \ref{distCEMP+CH}. 
   This is indicative that these stars have suffered from the external pollutions and hence changed their surface carbon abundances according to the depths of the surface convective zones. 
   CH stars may share a similar feature, but because of large metallicity, the giants can hardly display the strong feature of CH lines since the dilution of surface carbon by deepening surface convection may reduce the surface abundance below $\abra{C}{O} <1$. 
   In fact we see in Fig.~\ref{FeC} that CH giants are predominant in the lower metallicity while in the higher metallicity, abundant are Ba stars which are giant counterparts of CH stars with $\abra{C}{O} <1$.  

However, some differences are discernible between CEMP stars and CH stars. 
   The proportion of CEMP stars to EMP stars is significantly larger than that of CH stars to Pop II stars, as stated above. 
   CH stars and CEMP stars have different distributions with respect to the orbital separations, as shown in Figure \ref{OrbitD}, although the statistics may not be sufficient for CEMP stars with large separations. 
   In addition, some CEMP stars do not exhibit the enrichment of s-process elements, while all the CH stars show the fairly large enrichment. 
   The issue of these two subclasses, CEMP-s and CEMP-nos, is raised by \citet{Ryan05}. 
   There have been no explanations to where these features come from. 

The first purpose of this paper is to investigate the observational characteristics of CEMP stars by compiling data from the literature and to discuss their origins by applying the binary scenario, discussed above.
   The features of CEMP star population have been discussed only from the observational viewpoint by \citet{Lucatello05a} and by \citet{Ryan05}.  
   Relying upon the current understandings of stellar evolution, we present an interpretation on the origins of the both subclasses, CEMP-s and CEMP-nos and also of the stars enriched with nitrogen but with little carbon.   
   These subclasses are shown to be divided by their mass of primary stars.  

Secondly, we examine the binary scenario from the viewpoint of the evolution of binary systems with the mass transfer taken into account.  
   We corroborate our interpretation by confronting the theoretical predictions with the observed properties of CEMP stars and by analyzing the metallicity dependencies through the comparison with CH stars;  
    we deal with the relative frequencies, the correlations of orbital period and carbon-enrichment, and the distribution of evolutionary stages.

Finally, we inquire into the initial mass function (IMF) of EMP binaries on the basis of our binary scenario.  
   The observed properties of EMP stars are thought to give information on the formation history and early evolution of Galaxy.  
   From the viewpoint of chemical evolution, the richness of CEMP stars are linked with the IMF peaking at the intermediate stellar mass range \citep[e.g., $4-8 \msun$;][]{Abia01}. 
   On the other hand, \citet{Lucatello05b} discuss the observed frequency of CEMP-s stars from the binary scenario similar to ours, arguing an IMF, shifted slightly to higher mass as compared with those of metal-rich populations, such as expressed in a lognormal form peaking at $0.79\msun$.  
   These existent studies, however, seem not to properly take into account the recent progress in understanding the evolution of currently observed EMP stars, and in particular, the modifications of surface characteristics during their long lives.  
   \citet{Lucatello05b} ignore the differences in the evolutionary paths to the carbon enhancement across the metallicity of $\feoh \simeq -2.5$, theoretically predicted as stated above.  
   Furthermore, they deal only with CEMP-s stars, although it is desirable to put all other constituents of EMP stars into perspective as well.  
   In this work, we apply the current best knowledge on the evolution of metal-poor stars and consider not only CEMP-s stars but also with CEMP-nos stars and other subclasses.  
   Another important factor to the binary scenario is the assumption on the mass ratio distribution between the component stars, for which a consensus has not been reached observationally nor theoretically \citep[e.g., see][ and references therein]{Mazeh03}.   
   In this paper, we assume one of the plausible assumptions, though different from that assumed by \citet[][; see \S \ref{IMFsec}]{Lucatello05b}, and investigate the consequences.    

The initial mass function derived in this work will give a new insight into the nature of EMP stars and their roles in early evolution of our Galaxy and the formation of Galactic halo.  
   One proven feature of EMP stars is the scarcity as population in comparison with stars of more metal-rich populations. 
   Once \citet{Bond81} has raised an issue of ``G-dwarf'' problem in the Galactic halo on the basis of the fact that the number of stars decreases sharply with decreasing the metallicity below $\feoh \lesssim -1.7$ and no stars were found below $\feoh \lesssim -2.6$ at that time. 
   Thanks to the recent large-scaled HK and HES surveys, the known EMP stars have greatly augmented in number, and yet, it is true that they remain by far fewer as compared with other constituent, Pop II stars in the Galactic halo.  
   HES survey identifies $\sim 200$ stars of $\feoh < - 3.0$ in the magnitude range of $10 \lesssim B \lesssim 17.5$ among $\sim 40$\% of candidates selected from the fields of 8225 square degrees \citep{Beers05}.  
   Accordingly, EMP stars of $\feoh < - 3.0$ may occupy only a tiny fraction of stars, much fewer as compared with the stars in the globular clusters, ancillary constituents of Galactic halo.  
    We may also address this problem.  

This paper is organized as follows: 
   We start with reviewing the mechanisms to enhance the carbon abundance in the stellar envelope of EMP stars in \S 2. 
   Then we formulate our binary scenario in the following two sections;  the origin of abundance anomalies observed among EMP stars is identified in \S \ref{NandS} on the basis of the evolutionary models, and the parameters of binary systems is specified in \S \ref{binary} from the evolution of binary systems taking into account of the stellar wind from the primary and the subsequent accretion by the secondary.  
   In \S \ref{period}, the predictions of our binary hypothesis are confronted with the observations from particular CEMP stars on the period-carbon enhancement relations.   
   In \S \ref{statistics}, we investigate our binary scenario for CEMP-s stars, whose origin is well-defined theoretically and which deserve a separate statistical study because of their relative richness as compared with other EMP stars.  
   In \S\ref{IMFsec}, we deal with the both subclasses of CEMP-s and CEMP-nos stars to derive the IMF of EMP stars and to examine its implications. 
   Finally, \S \ref{conclude} concludes the paper.

\section{Evolutionary Characteristics of EMP Stars} 

   A basic distinction in the evolution between the EMP stars and the more metal-rich (Population I and II) stars is the engulfment of hydrogen by the helium convection during the helium flash in the core or in the shell.  
   Consequently, the low- and intermediate-mass EMP stars can enrich the surface with the nuclear products of helium burning through the helium-flash driven deep mixing (He-FDDM), triggered by the hydrogen mixing in addition to the third dredge-up (TDU) similarly in the metal-rich populations. 
   In this section, we review the evolution of EMP stars in relation to the carbon enhancement as well as to the s-process nucleosynthesis, and present a revision of the general picture of their evolution, originally formulated by \citet{Fujimoto00} as a functions of initial mass and initial metallicity. 

\subsection{Evolutionary Paths to Surface Carbon-Enhancement}

\citet{Fujimoto90} first show that the hydrogen mixing occurs during the helium flash in the core of a metal-free (Pop III) star of mass $1 \msun$ at the RGB tip, which is consequent upon a low entropy of the hydrogen burning shell, realized by low CNO abundances. 
   \citet{Hollowell90} compute the progress of subsequent He-FDDM on the RGB (He-FDDM-R) in detail and demonstrate that it enhances the surface abundance of carbon and nitrogen.  
   A number of following works obtain the consistent results of He-FDDM-R \citep{ Schlattl01, Schlattl02, Weiss04, Picardi04}.  
   This mechanism works for the stars of very low metallicity \citep[$\feoh \lesssim -4.5$;][]{Fujimoto95} and of low masses, $M \lesssim 1.1 M_{\odot}$ for $Z =0$ \citep{Suda04} and $M \lesssim 1.2 M_{\odot}$ for $\feoh \simeq -5.3$ (Suda 2006 in preparation). 

For the stars that do not undergo He-FDDM-R, \citet{Fujimoto00} show that the similar events happen during helium shell flashes in the early stage of thermal pulsating AGB phase.  
   He-FDDM on the AGB (He-FDDM-A) occurs in a star with larger metallicity ($\feoh \lesssim -2.5$) and of larger masses ($M \lesssim 3 M_{\odot}$). 
   \citet{Cassisi96} report the hydrogen mixing into helium convection during a helium shell flash for a star of $M = 0.8 \msun$ and $\log Z = -10$; 
   despite the sufficiently low metallicity, their model star has skipped He-FDDM-R presumably owing to the difference in the input physics affecting the evolution leading to the main He flash at the RGB tip.  
   The progress of He-FDDM-A is investigated by a few other authors \citep[][a $ 2 \msun$ star of $\feoh = -2.7$; Straniero et al.~2004, a $1.5 \msun$ star of $Z = 5 \times 10^{-5}$]{Iwamoto04} and all of these results are consistent with \citet{Fujimoto00}. 

\begin{figure}
\epsscale{1.0} \plotone{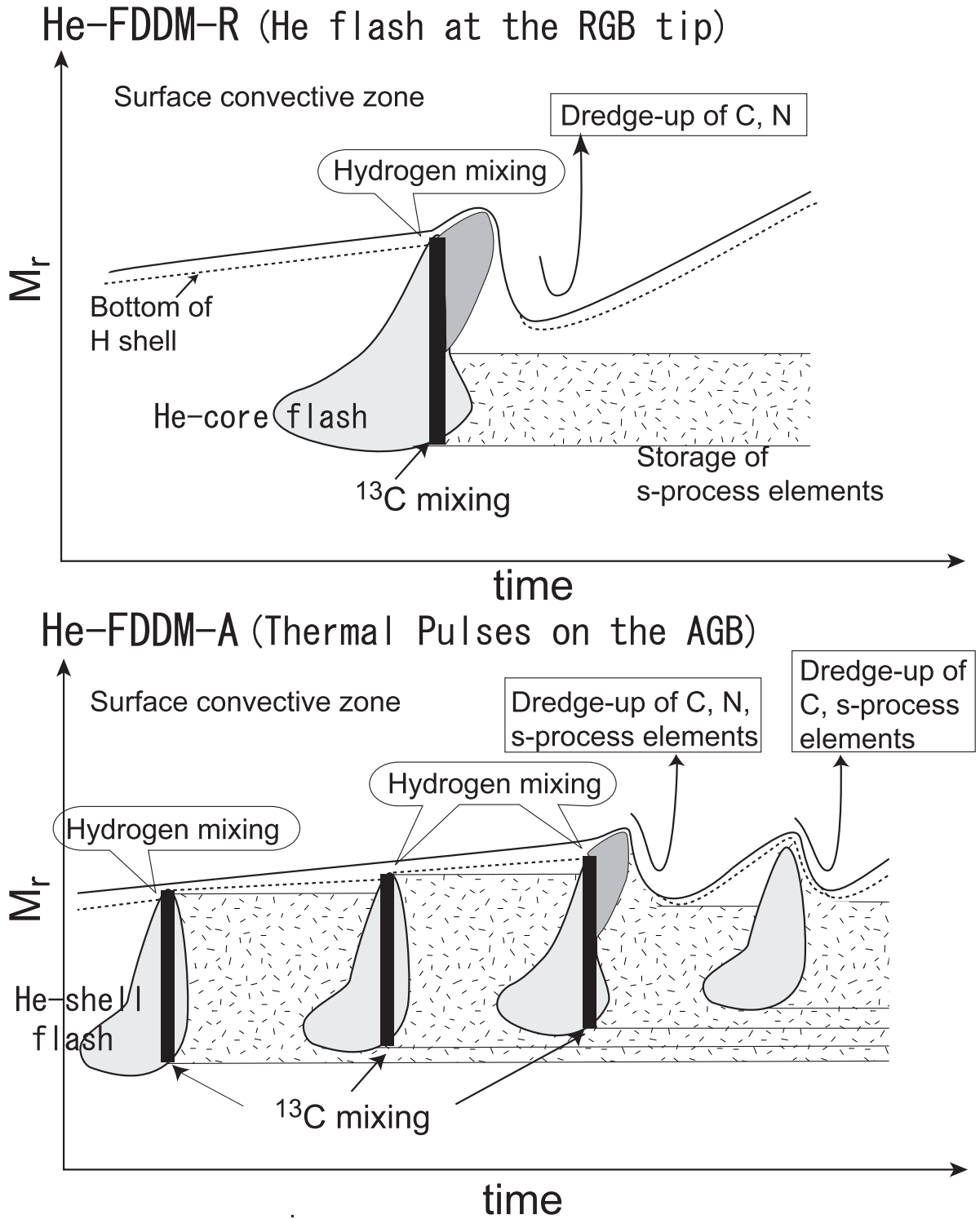}
\caption{Schematic drawings of the hydrogen mixing and progress of helium-flash driven deep mixing (He-FDDM): 
   The hydrogen mixing leading to splitting of the helium flash convective zone to trigger He-FDDM when the helium convection extends through the hydrogen containing layer for the first time (top panel): 
   Recurrent hydrogen mixings at small rates without splitting the helium flash convective zone, which ends with the strong mixing to trigger He-FDDM, and is followed by third dredge-up for stars of mass $M \gtrsim 1.5 M_\odot$ (bottom panel). }
\label{mixing}
\end{figure}

Figure~\ref{mixing} schematically illustrates the progress of He-FDDM during core flash on the RGB tip (upper panel) and He-FDDM during shell flashes on the AGB (lower panel).  
   The physical processes are quite similar, and yet, they may differ according to the rate of hydrogen mixing when the helium convection first makes contact with the hydrogen-containing layer \citep{Sweigart74}. 
   If the mixing rate of hydrogen is sufficiently large, the hydrogen engulfment causes the split of convective zone and triggers He-FDDM. 
   For the core helium flash, this is the case for the first flash because the core flash starts with a sufficient strength, as shown in Figure~\ref{mixing} (He-FDDM-R).  
   For the helium shell flashes on the AGB, on the contrary, the hydrogen mixing takes place while the shell flashes grow in strength along with the cooling of core, and hence, starts with weak one that cannot bring about the splitting of the helium convective zone.  
   After repeating the weak hydrogen mixing as it recurs, finally the shell flash grows strong enough to trigger He-FDDM, as shown in Figure~\ref{mixing} (He-FDDM-A).  
   During the He-FDDM(-R and -A), the mixed hydrogen is carried down and burns in the middle of helium convective zone and ignites a hydrogen shell flash to split the convection into lower and upper ones, driven by the helium and hydrogen burning, respectively. 
   In the decay phase of the hydrogen shell flash, the flash convection retreats in the upper zone and the shell occupied by the flash convection expands due to the heat deposited during the hydrogen flash and during the helium flash.  
   Consequent upon the shell expansion, the surface convective zone deepens in mass and penetrates down into the shells, formerly occupied by the hydrogen flash convection.
   As a result, the nuclear products of the helium flashes, processed by the hydrogen shell flash are bought up to the surface. 
   He-FDDM enriches the surface not only with carbon but also with nitrogen up to $\abra{N}{C} \lesssim 1$.  

In addition to He-FDDM, EMP stars undergo the third dredge-up (TDU) as the stars of metal-rich populations except for the very low metallicity.  
   There is a lower mass limit to the stars that undergo the third dredge-up; 
   \citet{Lattanzio86} shows that the TDU operates in a Pop II star of mass 1.5 $M_{\odot}$. 
   A trend is recognized that TDU occurs in less massive stars for lower metallicity \citep{Iben83a}, although the efficiency of dredge-up depends on the treatment of convection \citep[see][and references therein]{Karakas02}.  
   Furthermore, our low-mass models may no longer remain metal-poor as far as CN elements are concerned because of precedent He-FDDM.  
   Thus we may well set the lower mass limit for the TDU to occur at $\sim 1.5 M_{\odot}$, though this assumption has little to do with the following discussion.  
   This implies that stars in the middle mass range between $\sim 1.5 \msun$ and $\sim 3.5 \msun$ experience both, first He-FDDM and then TDU \citep{Iwamoto04, Straniero04}. 
   It is also known that in massive AGB stars, the temperature at the bottom of envelope convection reaches high enough to convert carbon into nitrogen and to reduce the $\abra{C}{O}$ ratio in the envelope even below unity \citep{Boothroyd93}, while the abundance of sum of CN elements increases as carbon is dredged up via TDU.  
   We take the lower mass limit to this hot bottom burning (HBB) to be $\sim 5M_{\odot}$, although it may depend on the metallicity and on the assumption about the efficiency of convective heat transport \citep[see, e.g.,][]{Ventura01}.  

\subsection{s-Process Nucleosynthesis in EMP Stars}

The hydrogen mixing may also promote the s-process nucleosynthesis in the helium convective zone.  
   The engulfed hydrogen is carried inward and captured by \nuc{12}C in the middle of convective zone and \nuc{13}C, thus produced, is mixed further inward to release neutrons through $\nucm{13}{C}(\alpha, n) \nucm{16}{O}$ reaction \citep{Iwamoto04,Suda04}. 
   Thus, He-FDDM brings about the surface enrichment with s-process elements, as illustrated in lower panel of Fig.~\ref{mixing}.  
   During the weak hydrogen mixing without the splitting of convective zone, the synthesized s-process elements are spread all over the helium convective zone, and will be involved again in the helium convective zones during the subsequent helium shell flashes. 
   Eventually as the shell flash grows strong enough to trigger He-FDDM-A, s-process elements are dredged up to the surface along with carbon and nitrogen. 
   For massive stars of $M \gtrsim 1.5 M_\odot$, a part of s-process elements, stored in the helium convective zone, are dredged up during the following TDU although their abundances suffer dilution by matter newly added to the helium core due to quiescent hydrogen shell burning, and then, incorporated into the flash convection.  
   On the other hand, He-FDDM-R may bring little s-process elements to the surface since the convective zone splits for the first hydrogen engulfment, as shown in upper panel, though some of \nuc{13}C is carried down into the lower helium convective zone before the splitting and gives rise to s-process nucleosynthesis.  

\begin{figure}
\epsscale{1.0} \plotone{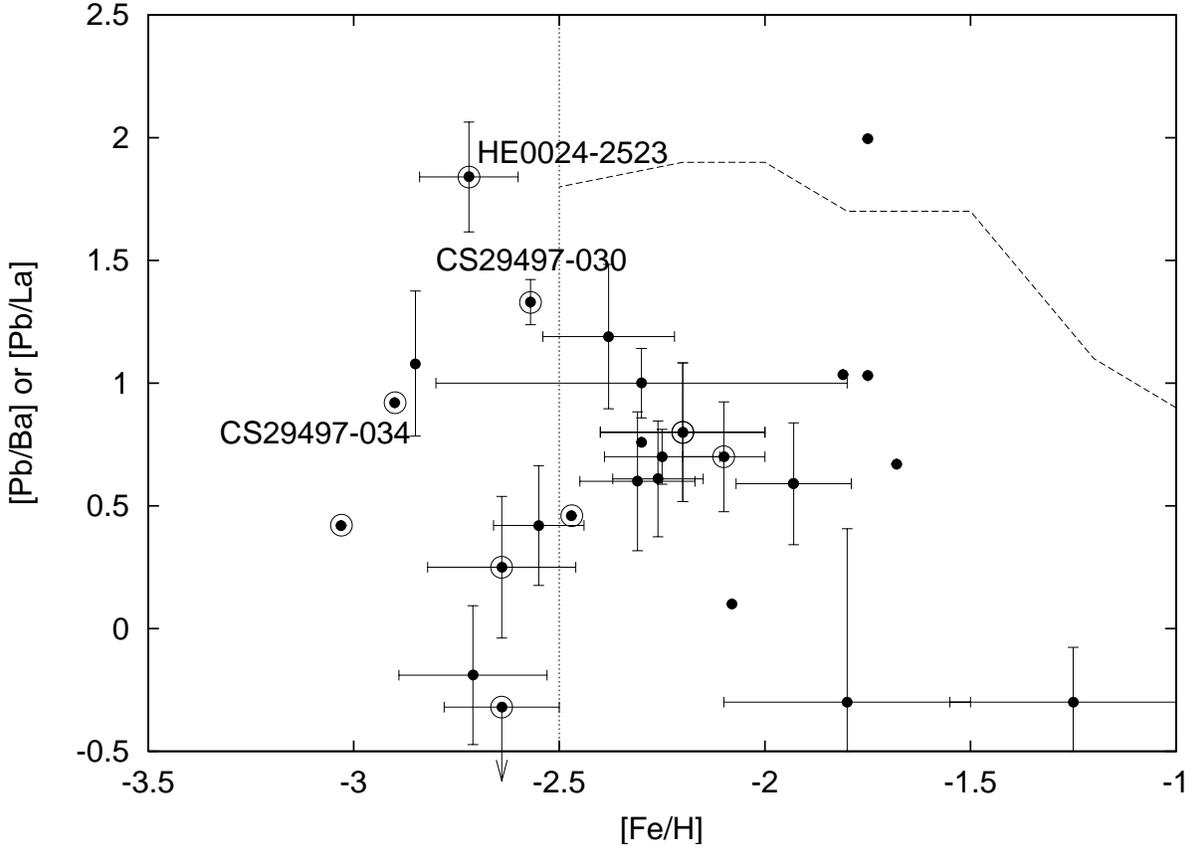}
\caption{Relative abundances between the main s-process elements Ba, or La, and the heavy s-process elements, Pb, plotted against the metallicity for CEMP and CH stars.  Double circles denote those with the orbital periods observed to date;  points with names attached denote three stars of the shortest periods, which are likely to have experienced common envelope evolution.  
Broken line denotes the prediction from the radiative \nuc{13}C burning model by \citet{Busso99}
}\label{PbBa}
\end{figure}

There has been proposed the other site of the s-process nucleosynthesis with the same reaction $\nucm{13}{C} (\alpha, n) \nucm{16}{O}$ as the neutron source but attendant on TDU in AGB stars; 
   it relies upon the alleged formation of thin layer containing \nuc{13}C, i.e., the \nuc{13}C pocket, in the top of helium zone during the dredge-up phase, presumably by semi-convection in the carbon-rich helium zone \citep{Iben82a, Iben82b} and/or by convective overshooting across the bottom of surface convection.  
   It is found that the \nuc{13}C in the \nuc{13}C pocket burns during the inter-pulse phase and the s-process nucleosynthesis occurs under the radiative condition before the subsequent helium shell flash is ignited \citep{Straniero95,Gallino98}.  
   This radiative \nuc{13}C burning model is argued to work for the metal-rich populations from comparison with solar relative abundances of s-process elements.  
   In particular, a tendency is proposed toward larger production of heavier elements for lower metallicity since the number of neutrons available per seed nucleus may increase with decreasing metallicity \citep[][for the review see Busso et al.\ 1999]{Busso95}. 
   From this radiative \nuc{13}C model, therefore, we may expect that EMP stars exhibit very large ratios between the heavy to main s-process elements, much larger than the stars of the metal-rich generations. 
   In actuality, however, the existent observations indicate that this is not necessarily true;  
   rather reported is an opposite trend of decreasing the $\abra{Pb}{Ba}$ ratio for smaller metallicity $\feoh \simeq -2.5$ \citep[][see also Lucatello et al.~2003]{Aoki00}.  
   Figure~\ref{PbBa} shows the ratios of heavy to main s-process elements, compiled from the literature, as a function of the metallicity. 
     A break in the variation of the $\abra{Pb}{Ba}$ ratio with metallicity is discernible near $\feoh \simeq -2.5$; 
   this is particularly evident if we note that two stars with the largest $\abra{Pb}{Ba}$ ratios are among the binaries with the three shortest periods, and hence, are likely to have been suffered from an extensive mixing during the common envelope evolution \citep{Lucatello03}. 

The above fact may be interpreted as evidence that the efficiency of \nuc{13}C pocket decreases and/or varies in EMP stars \citep[][Gallino et al, 2006, private communication]{Ryan01} or even as evidence that the radiative \nuc{13}C burning will not work for the metallicity below $\feoh \lesssim -2.5$ \citep{Suda04}.  
   The overshooting from the bottom of envelope convection may be conceivable in some way or other, but we have not yet a reliable theory for supporting the formation mechanism of \nuc{13}C pocket, and in particular, for predicting the variations with the metallicity.  
    In addition, for the metallicity of $\feoh \lesssim - 2.5$, the s-process nucleosynthesis occurs in the helium convective zone triggered by hydrogen mixing within the current standard framework of stellar structure, in which mixed \nuc{13}C is diluted over the entire helium convective zone and will not necessarily give such large numbers of neutrons allotted per seed nuclei as predicted from the radiative \nuc{13}C burning model \citep{Busso99}.  
    Accordingly we may well assume that there exists a critical metallicity near $\feoh \simeq - 2.5$ below which the radiative \nuc{13}C burning is ineffective during TDU. 
    This assumption will be further examined from the comparisons with the observations.  

\subsection{A General Picture of the Evolution to Carbon-Enhancement}

\begin{figure}
\epsscale{1.0} \plotone{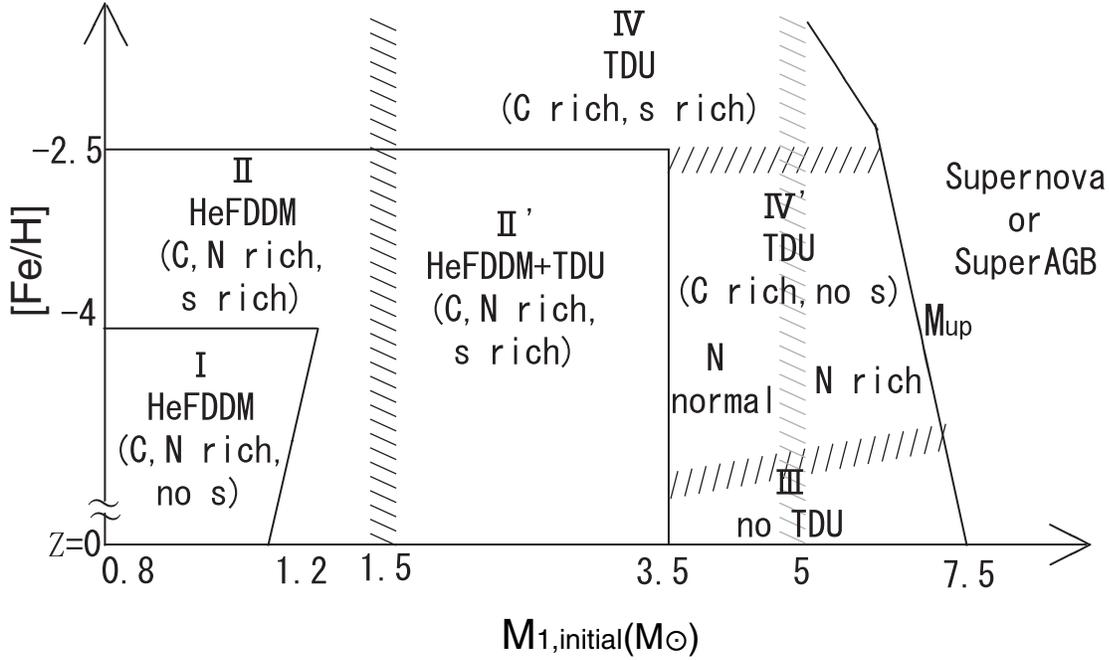}
\caption{A general picture of evolution of extremely metal-poor stars to the surface carbon-enhancement via helium-flash driven deep mixing (He-FDDM) and/or via third dredge-up (TDU) on the diagram of the initial mass $M_{1, \rm initial}$ and the pristine metallicity $\feoh$ with the CNO abundances in the solar ratio to Fe. 
   The most right-hand side is delineated by the upper mass limit, $M_{\rm up}$, to the development of carbon-oxygen, electron-degenerate core, taken from \citet{Cassisi93}, and hence, giving the lower mass limit to carbon ignition under electron non-degenerate condition. }\label{He-FDDM}
\end{figure}

The above discussion leads to a general picture of carbon and s-process enhancement in EMP stars, as illustrated in Figure \ref{He-FDDM}.  
   It is an up-to-date version of the dependences of the He-FDDM and TDU on the initial stellar mass and metallicity, formulated by \citet{Fujimoto00}.  

\begin{description}
\item[case I: ] Low-mass stars of $\feoh \lesssim -4.5$ in the mass range of $M \leq 1.1\,M_{\odot}$ for $Z = 0$ and $M\leq 1.2 \, M_{\odot}$ for $\feoh \simeq -5.3$ \citep[][Suda 2006 in preparation]{Suda04} undergo He-FDDM-R at the helium core flash. 
   Once He-FDDM-R occurs, the surface composition becomes such as $[{\rm C}/{\rm H}] \sim 0$ and $\abra{N }{ C} \sim 1$. 
   These stars may exhibit little enhancement of s-process elements. 

\item[case II and II$^{\prime}$: ] For stars located in the region of $\feoh \lesssim -2.5$ and $M \lesssim 3.5 \,M_{\odot}$ in Fig.~\ref{He-FDDM}, but excluding the region of case I, He-FDDM-A occurs when they evolve into the early phase of thermally pulsing AGB (TP-AGB). 
    The surface abundance of carbon reaches as large as $[\abra{C}{H}] \sim 0$ for the stars of the smallest masses, while it decreases for the stars of larger masses because of larger envelope mass and also of smaller mass in the flash convective zone. 
   He-FDDM-A ceases to occur after the CN abundances in the envelope exceeds $[{\rm C}+ {\rm N} / {\rm H}] \simeq -2.5$, and the subsequent evolution is divided by the occurrence of TDU; 
   case II of $M < 1.5 \msun$ that no further mixing takes place and case II$^{\prime}$ of $M \gtrsim 1.5 \msun$ that TDU follows and brings C synthesized by helium burning to the surface to reduce the N/C ratio. 
   S-process elements, synthesized via the convective \nuc{13}C burning, are dredged up by He-FDDM-A for the both cases, and in addition, by the subsequent TDU for case II$^{\prime}$.  

\item[case III: ] The stars more massive than $\sim 3.5 \msun$ and with the smallest metallicity will undergo neither He-FDDM nor TDU, although the latter point is a matter of controversy because of the efficiency of overshooting \citep{Suda04}.  

\item[case IV and IV$'$: ] The stars in these region undergo TDU but not He-FDDM.  Cases IV and IV$'$ are distinguished by the surface enrichment of s-process elements, i.e., case IV$'$ of lower metallicity without s-process enhancement and case IV of higher metallicity with s-process enhancement. 
   The difference is ascribed to the efficiency of the radiative \nuc{13}C burning.  
   TDU dredges up carbon, but it may be converted into nitrogen via the hot bottom burning in the envelope for massive stars of $M\gtrsim 5 \msun$; 
   thus, the nitrogen abundance may vary with the stellar mass from no enhancement to the large enhancement even exceeding the carbon abundance and up to the equilibrium value of CN (or CNO) cycles ($\abra{N}{C} = 30 \sim 16$ in the range of $T = 5 \cdot 10^7 \sim 10^8$ K). 

\end{description}

In summary, the surface carbon enhancement in EMP stars is brought about by He-FDDM for the initial mass $M \lesssim 1.5\msun$, by both He-FDDM and TDU for $1.5 \msun \lesssim M \lesssim3.5\msun$, and only by TDU for $M \gtrsim 3.5\msun$. 
   One of the distinguishing features of He-FDDM, as compared with TDU, is the surface enrichment of nitrogen at the same time with that of carbon; 
   for cases I and II, the surface abundance ratio results ${\rm N}/{\rm C}\simeq 1 - 1/5$ \citep{Fujimoto00}, while for case II$'$, it decreases as carbon is enriched by TDU.  
   For case IV$'$, the nitrogen abundance may range from no enhancement to a large value corresponding to the equilibrium in CN (or CNO) cycles. 
    As for s-process elements, our scenario predicts the surface enrichment due to the convective \nuc{13}C burning for cases II and II$'$, and postulates no or little enhancement for case IV$'$.  
   This conjecture predicts the variations in the abundances of nitrogen and of s-process elements in CEMP stars as well. 

\section{Binary Scenario I  ---  The Origins of CEMP and Related EMP Stars}\label{NandS}

If EMP stars belong to a close binary system, the surface characteristic abundances that the evolved primary star has developed can be imprinted onto the secondary star through the mass transfer.  
   A CEMP star, thus produced, may keep a record of the primary star on their surface characteristics.  
   In this section, we survey the observed characteristics of CEMP stars and other related EMP stars, and discuss their origins, in the light of the modifications of surface abundances of carbon, nitrogen and of s-process elements, as described in the preceding section.   

We compile the characteristics of CEMP and CH stars from the literature as summarized in tables \ref{table} and \ref{CHtable}, respectively.  
   Our criterion of CEMP stars is that $\feoh \le -2.5$ and $[\abra{C}{Fe}] \ge 0.5$, but we include those with slightly larger metallicity which show the nitrogen enhancement ($[\abra{N}{Fe}] \ge 0.5$ ) with taking into account possible errors ($\sim 0.25$ dex) in the abundance determination.  
   We have 34 CEMP stars in total (see Fig.~\ref{FeC}), list their available data of the orbital parameter, the gravity, the metallicity, the abundances of C, N, O and s-process elements (Sr, Ba, Pb) relative to iron; 
   7 stars have the orbital periods derived with the maximum $P = 11.3$ yr and 7 more stars are suspected of binarity from variations of radial velocity.  
   As for CH stars, we compile 85 stars by the criterion that $\feoh > -2.5$; in addition to those classified as CH stars in literature, we include three blue stragglers showing strong CH lines, and also, Ba stars which are thought as giant counterparts of subgiant CH stars with $\abra{C}{O} <1$ \citep[e.g.,][]{Luck91}. 
   38 stars of them have the orbital periods determined (two stars with the lower bounds).  
   Note that CH stars are divided into two groups of CH (giant) stars  and subgiantCH stars, the latter of which includes both dwarfs and subgiants \citep{Bond74}.  

\subsection{Sub-classification of CEMP Stars}

Figure~\ref{BaFeNCFe} shows the observed relationship of the enrichment between Ba and nitrogen among CEMP stars.  
   As already noticed by \citet[][see also Ryan et al.~2005]{Aoki02a}, they are rather clearly divided into two groups of $[\abra{Ba}{Fe}] \gtrsim 1$ and $[\abra{Ba}{Fe}] \lesssim 0$, where a few exceptions include CS22892-052 enriched with r-process elements \citep{Sneden96}. 

All CEMP stars with Ba enrichment ($[\abra{Ba}{Fe}] \gtrsim 1$), which are defined as CEMP-s, are enriched with nitrogen ($[{\rm N}/{\rm Fe}] \gtrsim 1$). 
   The relative abundance to carbon lies in the range of $0.01 \lesssim {\rm N}/{\rm C} \lesssim 1$, and is consistent with the theoretical predictions of cases II and II$'$; 
   the largest ratios may result from He-FDDM in case II and the smaller ratios are explicable in terms of the increase of carbon via TDU in case II$'$. 
   
On the other hand, CEMP stars without Ba enrichment ($[\abra{Ba}{Fe}] < 0$), which are defined as CEMP-nos, exhibit still larger variations in the nitrogen abundances, ranging from no enhancement $[\abra{N}{Fe}] \simeq 0$ (CS 22877-001) to much larger enhancement with the N/C ratios $ \simeq 10$ (CS22949-037). 
   The range of enrichment well fits to case IV$'$ evolution in our scenario with and without hot bottom burning. 
   It is difficult to search for the site of nucleosynthesis in the stars other than hot bottom burning that can realize such large ${\rm N}/{\rm C}$ ratios as near to the equilibrium values in the CN cycles. 
   As a class, CEMP-nos stars have the carbon enhancement smaller than CEMP-s stars, as pointed out by \citet[][also see Fig.~\ref{BaFeNCFe}]{Aoki02a}.  
   This is attributable to the dilution of carbon, dredged up, in the envelope of the primary star, which is more massive in case IV$^\prime$ than in cases II and II$^\prime$, and also, to the transformation of carbon into nitrogen by hot bottom burning.  

Some of stars in the CEMP-nos subclass have large oxygen abundance, comparable to, or even larger than, the carbon abundance.   
   In the binary scenario, the enhancement of oxygen exceeding about a tenth of carbon abundance has to be attributed to the pristine abundance \citep[cf. ][]{Suda04}, which may well be explained in terms of the general tendency of $[\abra{O}{Fe}]$ increasing with decreasing metallicity and its extension to EMP stars \citep[][but see also Garcia Perez et al. 2005]{Israelian98,Boesgaard99} (see appendix A.1). 

Among EMP stars, a class of stars display large nitrogen enrichment with $[{\rm C} + {\rm N} / {\rm Fe}] > 0$ but display only weak or little carbon enhancement with the ratio $\abra{C}{O} < 1$, which are classified as ``mixed'' stars by \citet{Spite05}. 
   We propose that these stars are made by the same mechanism as CEMP-nos stars but carbon was converted into nitrogen by hot bottom burning in more massive primary stars than that of CEMP-nos stars (appendix A.2). 
   In our compilation, those with r-process element enhancement such as CS22892-052 may also be included in the subclass of CEMP-nos since the Ba enrichment can be attributed to the r-process nucleosynthesis \citep{Sneden96}.  
   We will inquire their origin separately in this paper (\S \ref{NatureEMP} and appendix A.3). 

In our binary scenario, the origins of subclasses of EMP stars are divided according to the mass of their primary stars. 
  CEMP stars enriched with s-process elements (CEMP-s) are formed in the binary system with a primary star of mass in cases II and II$^\prime$ while CEMP stars without the s-process element enrichment (CEMP-nos) in the binary system with a primary of mass in case IV$^\prime$.  
   The nitrogen-enhanced ``mixed'' stars may be assigned to the binary systems, same as CEMP-nos stars, but with more massive primaries.  

\section{Binary Scenario II  ---  Binary Parameters of CEMP Stars } \label{binary} 

In this section, we discuss the binary parameters that the secondary star can evolve to CEMP stars through the mass transfer. 
   The first condition is that the primary star is allowed to develop the surface carbon enhancement. 
   The second one is that the binary system can enrich the secondary star sufficiently with the envelope mass ejected from the primary stars.  
   These conditions set lower and upper bounds to the binary separations, respectively, for a given set of the masses of primary and secondary stars, as discussed below.    
   We can derive the relation of the carbon abundance of secondary star to the mass of primary stars and the orbital separations (periods), which may be compared with the observations. 

\subsection{Roche Lobe Overflow} 

We start with the discussion of the first condition.  
   In close binary systems, the primary star expands as it evolves, and sometimes, fills its Roche lobe to lose the envelope through the mass transfer to the secondary and/or the mass out-flow from the systems \citep[e.g., ][]{Iben85}.  
   If this Roche lobe overflow (RLO) happens before the star becomes a carbon-enhanced star, then this system cannot yield a CEMP or CH star. 
   In order to yield a CEMP-s star, therefore, the initial size of the Roche lobe has to be greater than the stellar size when the primary star for $M \lesssim 3.5 \msun$ starts the He-FDDM.  
   The same is true for a CEMP-nos star and for a CH star but with the stellar size when TDU starts. 
   Since the volume of Roche lobe is given as a function of the mass ratio $M_2/M_1$ and the separation $A$ \citep{Paczynski71}, the lower bounds, $A_{\rm He-FDDM}$ and $A_{\rm TDU}$, to the initial binary separation necessary to accommodate the primary stars when He-FDDM and TDU start, may be derived from the radii, $R_{\rm He-FDDM}$ and $R_{\rm TDU}$, at the stage of He-FDDM and TDU as; 
\begin{eqnarray}
A_{\rm He-FDDM} ({\rm or} \ A_{\rm TDU}) & = & \frac{R_{\rm He-FDDM} ({\rm or} \ R_{\rm TDU})}{  0.38 + 0.2 \log (M_1/M_2)}.
\end{eqnarray}
   Figure \ref{dependences} shows these lower bounds on the diagram of the initial stellar mass of primary star and the binary separation for the stellar models taken from existent computations \citep[][, Suda 2006 in preparation]{Iben75, Lattanzio86, Marigo02,Suda04}. 
   The initial orbital radius needs to be greater than $0.2 \sim 1$ AU for a CEMP-s star while greater than $\sim 1$ AU is necessary for a CH star and for a CEMP-nos star.  

After the primary has developed the surface carbon-enrichment, the RLO may occur on AGB before the primary evolves to a white dwarf.  
   In this case, the binary will undergo common envelope evolution and shrink in size to be a short-period system or even to coalesce in some cases. 
   These systems may be distinguished from CH and CEMP stars produced through the accretion process in the wind from the primary star in the binary of wider initial separation as proposed by \citet{Boffin88} and discussed in the following. 

\subsection{Wind Accretion from AGB Stars}\label{wa}
 
The second condition, i.e., an upper bound to the binary separation, results from the wind accretion.  
AGB stars in binary lose their envelope through stellar wind \citep{Reimers75,Gail87} if the separation are sufficiently large to avoid RLOs. 
   Then, the secondary stars may accrete carbon-rich material in this wind to be carbon-enhanced stars. 
   The process for the secondary stars to accumulate the gas from the wind may well be approximated to the Bondi accretion \citep{ Bondi44}, and the cross section, $\sigma_{\mathrm{acc}}$, may be evaluated from the following formula, 
\begin{equation}\label{Bondi}
\sigma_{\mathrm{acc}} = \pi \left( \frac{2GM_{2}}{v_{\mathrm{rel}}^{2}}\right)^2, 
\end{equation} 
   where $G$ is the gravitational constant, $M_{2}$ the mass of the secondary star, and $v_\mathrm{rel}$ the velocity of the wind relative to the secondary. 
   Thus the mass accretion rate $d M_{2}/d t$ of the secondary star is related to the mass loss rate $d M_{1}/d t$ of the primary as
\begin{equation} 
\frac{d M_{2} (t)}{d t} = -\frac{G^{2} M_{2} (t)^{2}}{A(t)^{2} v_{\mathrm{rel}} (t)^{4}} \frac{v_{\rm rel} (t)}{v_{\rm wind}} \times \frac{d M_{1} (t)}{d t}, \label{eq:m1m2} 
\end{equation} 
   where the separation $A(t)$ between the two stars has been introduced as a function of time, $t$, and $v_{\rm wind}$ is the wind velocity with respect to the primary star.  The relative speed of the secondary star to the stellar wind is given by
\begin{equation}
 v_{\mathrm{rel}} = \sqrt{ v_{\mathrm{wind}}^{2}+\left|\vec{v}_{1} - \vec{v}_{2}\right|^{2} } \label{eq:vreldef} 
\end{equation} 
   with the orbital velocities, $\vec{v}_{1}$ and $ \vec{v}_{2}$ of primary and secondary stars.  The mass accreted by the secondary star depends on the separation $A(t)$ and the relative velocity $v_{\mathrm{rel}} (t)$, which may vary during the mass transfer along with the masses  $M_{1}(t)$ and $M_{2}(t)$.  

Suppose that the primary and secondary stars rotate on each circular orbit around the center of mass while the primary loses mass at a constant rate $d {M}_{1} / d t$, then the basic equations governing the evolution of a binary system are
\begin{eqnarray}
& & M_{1}(t) {v_{1}(t)^{2}} / {r_{1}(t)} = M_{2}(t) {v_{2}^{2}(t)} / {r_{2}(t)} \nonumber \\
& & \phantom{ M_{1}(t) {v_{1}(t)^{2}} / {r_{1}(t)}} = {GM_{1}(t) M_{2}(t)} / {A^{2}(t)}, \\
& & M_{1}(t)r_{1}(t) - M_{2}(t)r_{2}(t) = 0, \\
& & r_{1}(t) + r_{2}(t) = A(t).
\end{eqnarray} 
   Here $r_{i}$ denotes the radius of the orbit of star $i$ and $v_{i}$ the velocity. 
   The subscript $i=1$ refers to the values of the primary star and $i=2$ to those of the secondary star. The timescale for the change of $A(t)$ is assumed to be much longer than the orbital period.

If Jeans' theorem is applied, the relation between the separation and the mass of the binary system is given by 
\begin{eqnarray} 
A(t) [M_{1}(t) + M_{2}(t)] =  \lambda, 
\label{eq:sepmass}
\end{eqnarray} 
   where $\lambda$ is a constant, given by the initial conditions at time $t_{\mathrm{initial}}$. 
   This implies that the wind is assumed to carry the angular momentum away from the system without transferring it to the other material.  
   By expressing $A(t)$ and $v_\mathrm{rel} (t)$ in terms of the stellar masses and the wind velocity, equation (\ref{eq:m1m2}) is reduced to
\begin{eqnarray}
\frac{d M_{2} (t)}{d M_1 (t)}& = & - \left[ {M_2 (t) \over M_1 (t) + M_2 (t) } \right]^2 \left[ { \mu \over M_1 (t) + M_2 (t)} \right]^{-1} \nonumber \\
& & \quad \times \left[ 1 + { \mu^2 \over \left( M_1 (t) + M_2 (t) \right)^2 } \right]^{-3/2},
\label{eq:m2om1}
\end{eqnarray} 
   where $\mu$ is a constant, defined as
\begin{equation}
\mu \equiv \sqrt{{\lambda v_{\rm wind}^2}/{G}} = \sqrt{ 2 M_1 (M_1 + M_2) (A / R_1)} \left[ v_{\rm wind} / v_{\rm esc}  \right].  
\end{equation} 
   In the most rightmost member, we normalize the wind velocity with respect to the escape velocity, $v_{\rm esc}$, from the surface of the primary star $ (= \sqrt{2 G M_1 / R_1}$, where $R_1$ is the radius of primary star). 

If $A \gg R_1$, then, $\mu /(M_1 + M_2) \gg 1$ (since $v_{\rm wind} \sim v_{\rm esc}$) so that the variation of $M_2$ is much smaller than that of $M_1$. 
   We may neglect the variation of $M_2$ in the right-hand side and integrate eq.~(\ref{eq:m2om1}) until the primary becomes a white dwarf at time $t_{\rm final}$ to find the accreted mass, $M_{\mathrm{acc}}$, on the secondary star; 
\begin{eqnarray}
M_{\mathrm{acc}}  = & &
\left[ \frac{M_2^2}{ \mu } \left(\frac{M_1+M_2} {\sqrt{ \mu ^2 + (M_1 + M_2)^2}} - \log \left( \frac{1}{\mu } \left( M_1  + M_2  \right. \right. \right. \right.\nonumber \\ 
& & \left. \left. \left. \left. + \sqrt{\mu ^2 + (M_1 + M_2)^2 } \right) \right) \right) \right] ^{M_1 = M_1 (t_{\rm final})} _{M_1 = M_1 (t_{\rm initial}).} \label{eq:macc2-2}
\end{eqnarray} 
   The relation between the initial stellar mass, $M_1 (t_{\rm initial})$, and the mass, $M_1 (t_{\rm final})$, of C+ O core for metal-poor stars may well be approximated to 
\begin{eqnarray}\label{eq:coremass}
M_1 (t_{\rm final}) & & =  \max [0.54 + 0.073  M_1 (t_{\rm initial}), \min[0.29  \nonumber \\ 
& & + 0.178 M_1 (t_{\rm initial}), 0.65 + 0.062 M_1(t_{\rm initial})]],
\end{eqnarray} 
 that for the metallicity of $Z_0=0.001$, given by \citet{Han95}.

\subsection{Dilution due to Surface Convection }\label{cad}

Suppose that the matter of constant carbon abundance, $X_{\rm C, p}$, is accreted onto the secondary of the initial carbon abundance $X_{\rm C, 2, 0}$. 
   If the mass in the surface convection, $ M_{\rm scz, D}$, remains constant through mass accretion, then, the variation of surface carbon abundance, $X_{\rm C, 2}$, of the secondary star may be described by;  
\begin{equation} 
\frac{dX_{\rm C,2 }}{d t} = \frac{ X_{\rm C, p } - X_{\rm C, 2 } }{M_{\rm scz, D}} \frac{d M_2}{d t} . 
\end{equation} 
   The integration yields $X_{\rm C, 2}$ as a function of the accreted mass, $ M_{\rm acc}$, as 
\begin{equation} 
X_{\rm C, 2}=X_{\rm C, 2 ,0} + \left( X_{\rm C, p } - X_{\rm C, 2, 0} \right) \left[1 - \exp ( - M_{\rm acc}/M_{\rm scz, D}) \right]. 
\label{CenhanceD}
\end{equation}

The mass transfer is likely to occur while the secondary star resides on the main sequence with a shallow surface convective zone. 
   Then, the surface abundance suffers an additional dilution owing to the deepening of surface convection as the secondary later evolves to the red giant branch.  
   For the metallicity of $\feoh \lesssim -3$, a star of mass $M_{2} \simeq 0.8 \msun$ has the surface convection of mass $M_{\mathrm{scz, D}} \simeq 0.003\,M_\odot$ on the main sequence. 
   The surface convection, though once decreasing as the hydrogen abundance decreases in the center, turns to deepen rapidly as the shell burning starts and attains at the deepest reach of $M_{\mathrm{scz, G}} \simeq 0.20 \sim 0.35\,M_\odot$ in mass near the base of red giant branch when the luminosity $L \simeq \hbox{several} \times 10 \ L_\odot$ \citep{Fujimoto95,Suda04}. 
   At the base of RGB, therefore, the surface carbon abundance reduces to be 
\begin{equation}
X_{\rm C, 2} = X_{\rm C, 2, 0}(1 - M_{\rm acc}/M_{\rm scz, G}) + X_{\rm C, p} ( M_{\rm acc}/M_{\rm scz, G})  
\label{CenhanceG}
\end{equation}
   (here we assume $M_{\rm scz, G} > M_{\rm acc} + M_{\rm scz, D}$). 
   After that, it remains constant along the giant branch.  
   As a corollary, if carbon is brought by external pollution, the $[\abra{C}{H}]$ value decreases up to the maximum of 2 dex when an EMP star evolves to a giant. 

The surface carbon abundance of secondary stars resulting from the wind accretion is given as a function of the initial masses of component stars and the initial binary separation from eq.~(\ref{eq:macc2-2}) and from eq.~(\ref{CenhanceD}) or eq.~(\ref{CenhanceG}) for dwarfs and giants, respectively.   
   These relations may be transferred to the upper bound, $A_{M}(M_1, M_2, {\rm [C/H]}, M_{\rm scz})$, to the initial binary separation that allows the binary system with the primary star of mass $M_1$ to enrich the secondary star of mass $M_2$ that has the surface convection of mass $M_{\rm scz}$ with carbon above $[\abra{C}{H}]$ through the wind accretion.  
   In Fig.~\ref{dependences}, we plot the upper bounds of initial binary separation that give the carbon enhancement of $[\abra{C}{H}] = -1$, $-2$, and $-3$ for the secondary of giants with the surface convection $M_{\rm scz} = 0.35 \msun$ ($M_2 = 0.8 \msun$).  
   The carbon enhancement necessary for CEMP and CH stars may be taken to be $[\abra{C}{H}] > -3$ and $-1$, respectively.  
   For such a large initial separation, since the orbital velocity is much smaller than the wind velocity, the cross section of Bondi-Hoyle accretion reduce to a constant, $\sigma_{\rm acc} = \pi (G M_2 /v_{\rm wind})^2$, and hence, we may approximate   
\begin{eqnarray}
A_{M}(M_1, \,{\rm [C/H]},\, M_{\rm scz}) =  \left[ \{M_1-M_{1} (t_{\rm final}) \} \right. & &\nonumber \\
\times \sigma_{\rm acc} / (4 \pi 10^{\rm [C/H]-[C/H]_p} \left. M_{\rm scz} ) \right]^{1/2}.  & & 
\label{am}
\end{eqnarray} 
   The upper bound of separation increases in inversely proportion nearly to the second power of the carbon enrichment of secondary star, and hence, CEMP stars may have much wider separations than CH stars.  
   The bound also depends on the carbon abundance, $[\abra{C}{H}]_{\rm p}$, in the wind.  
   For TDU, the latter is currently subject to uncertainties for lack of the reliable theory on the mass loss, and yet, we may well assume that $[\abra{C}{H}]_{\rm p} \simeq 0$ or so. 

Now we have formulated our binary scenario as a function of the mass ratios and the binary separations, as illustrated in Fig.~\ref{dependences}.  
   In the following three sections, we examine the validity of our interpretation and binary scenario firstly as to the relationship between the orbital period and carbon enhancement, secondary as to the statistical properties of CEMP-s stars, and finally, as to the relative frequencies of CEMP-s and CEMP-nos stars and discuss the initial mass function of EMP stars.  

\section{Period-Carbon Enhancement Relation}\label{period}

We can deduce the relations between the carbon abundance of the secondary star and the orbital period from our modeling in the preceding section. 
   The accreted mass, $M_{\rm acc}$, onto the secondary star can be estimated from eq.~(\ref{eq:macc2-2}) as a function of the initial masses of component stars, $M_1 (t_{\rm initial})$ and $M_2 (t_{\rm initial})$ and the initial separation, $A (t_{\rm initial})$, and then, the surface chemical compositions of the secondary stars may be evaluated with the mixing in the surface convective zone of the secondary star taken into account from eq.~(\ref{CenhanceD}) or eq.~(\ref{CenhanceG}). 
   On the other hand, as a kinematic feature, the orbital period after the primary star becomes a white dwarf is calculated from the final separation $A(t_{\rm final})$ in eq.~(\ref{eq:sepmass}) and the final mass $M_1(t_{\rm final})$ of the primary star in eq.~(\ref{eq:coremass}). 
   These are observable and can be used to test the present scenario, as in the similar way to \citet{Boffin94} who discuss the relationship between the overabundances of s-process elements and the orbital periods for barium stars. 

In deriving the relationship between the accreted mass and the orbital period, we have applied Jeans' theorem. 
   In actuality, however, the wind material is thought to carry an additional orbital angular momentum unless the wind velocity $v_{\rm wind}$ is much larger than the orbital velocity $v_{\rm orb} \ (= \vert \vec{v}_1 - \vec{v}_2 \vert)$. 
   In fact, hydrodynamical calculations of wind accretion \citep[e.g.,][]{Nagae04, Jahanara05} have shown that the orbit shrinks if $v_{\rm wind} \lesssim (1.5 \sim 1.7) v_{\rm orb}$ for the mass ratio of $q (\equiv M_2/M_1) = 1 \sim 1/3$ \citep[see also][]{Hachisu99}, which may correspond in our scenario to; 
\begin{eqnarray}
A & \lesssim & 5 \sim 7 \ (1 + q) ({M_1/M_\odot}) (v_{\rm wind}/ 20 \hbox{ km s}^{-1})^{-2} \hbox{ (AU)}. 
\\
P & \lesssim & 11 \sim  16 \ (1 + q) ({M_1/M_\odot}) (v_{\rm wind}/ 20 \hbox{ km s}^{-1})^{-3} \hbox{ (yr)}. 
\label{binary-shrinkS} \label{binary-shrinkP}
\end{eqnarray} 
   Jeans' theorem holds good for the wider binaries of $ v_{\rm wind} \gtrsim  (1.8 \sim 1.9) v_{\rm orb}$, while the flow structure resembles the Roche lobe over-flow (RLO) for the wind velocity of $v_{\rm wind} < 0.4 \ v_{\rm orb}$.  

\begin{figure}
\epsscale{0.7} \includegraphics{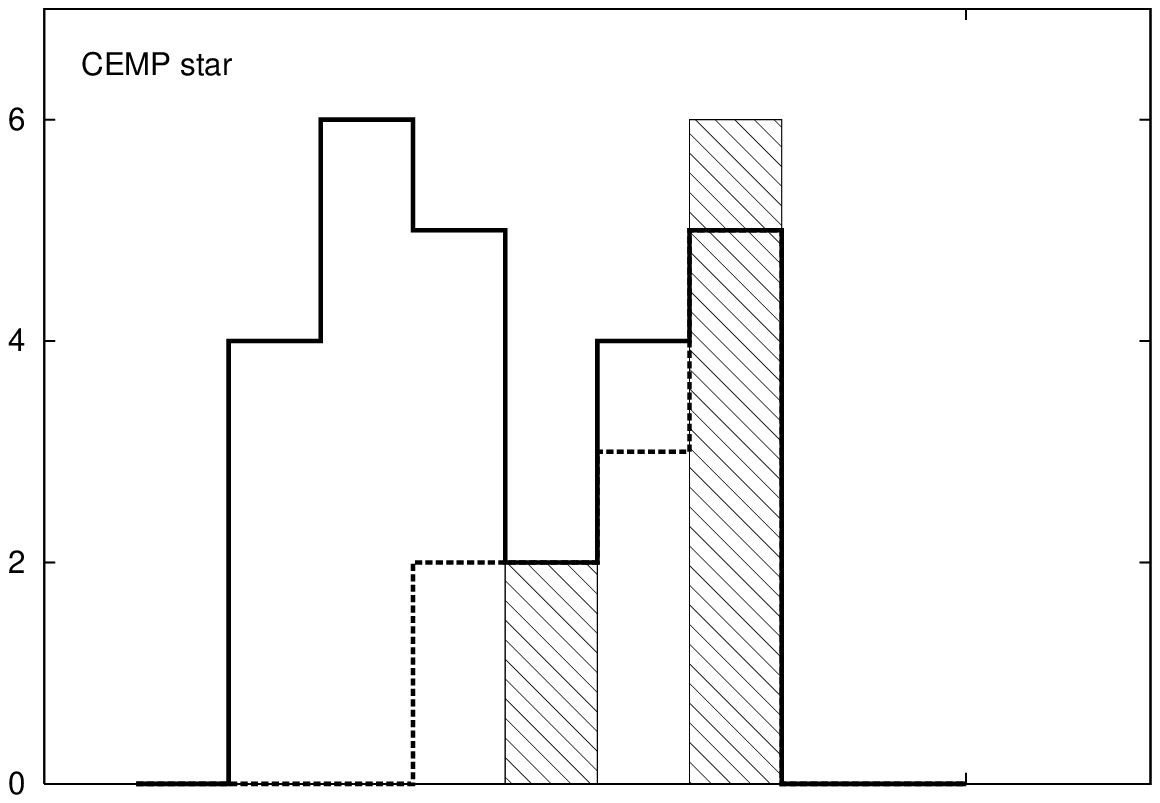} \includegraphics{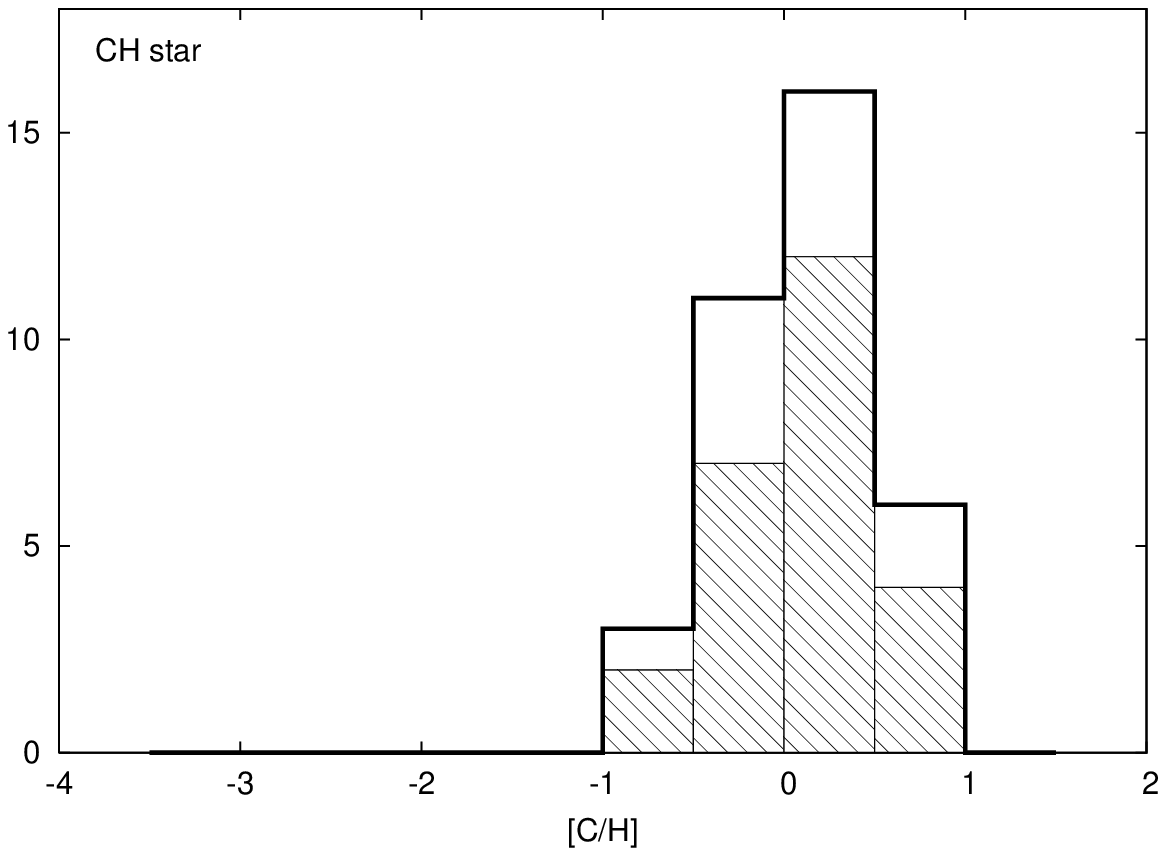}
\caption{Distributions of carbon abundance, $[\abra{C}{H}]$, among giants (columns enclosed by thick lines) and dwarfs + subgiants (shaded columns) of CEMP stars (upper panel) and of CH 
 stars (lower panel);  in upper panel, CEMP-s + CEMP-nos and CEMP-s giants are enclosed by solid and broken lines, respectively. 
   CEMP giants spread in a wide range down to low $[\abra{C}{H}]$ values because of dilution by a factor of up to $\sim 100$, a factor corresponding to the difference in the depth of surface convection of giants and dwarfs.  
   For CH giants, the distribution is rather narrower since the surface carbon abundance is diluted to be ${\rm C}/{\rm O} <1$ because of large pristine abundance of oxygen. 
   The spreads in the surface carbon abundance among dwarfs and subgiants may be attributed partly to the difference in the carbon enrichment in the donor primary stars and partly to the dilution by surface convection.  
} \label{distCEMP+CH}
\end{figure}

Figure \ref{ch-orbit} plots the relations between the surface carbon abundance $[\abra{C}{H}]$ of secondary stars and the orbital period, calculated under the assumption of Jeans' theorem with the primary stars of mass 1.5 and $3 \msun$.  
   The bottom panel is for the binary systems of $\feoh > -2.5$ whose primary stars have undergone the carbon enhancement by TDU, and may be compared with the observations for CH stars. 
   Here we assume the carbon abundance in the wind matter to be $[\abra{C}{H}]_{\rm \ p} = 0.5$ and the wind velocity to be $v_{\rm wind} = 20 \ \hbox{km}s^{-1}$.  
   A subgiant has thinner surface convective zone (by a factor of 30 in mass) than a red-giant star, which yields larger carbon enhancement with the same binary parameters.  
   For the systems of short period $P \lesssim \hbox{a few} \times 10$ yr, the subgiants tend to have the same surface abundances as the wind.  
   On the other hand, the giants display the carbon abundance smaller than the wind even for the smallest separations, and decreases for larger separations nearly as $[\abra{C}{Fe}] \sim -4/3 \log P$.  
   These tendencies agree with the observations that the subgiants tend to have larger carbon abundance than the giants; 
   the exceptions are the blue stragglers of smallest metallicity of $\feoh <-2$, which may be attributed to the difference in the carbon abundances reached in the primary stars, as discussed below.   
   For most of subgiant CH stars, the observed carbon abundances are larger than the solar value, which indicates that the carbon abundance in the wind is above the solar value (up to $[\abra{C}{H}]_{\rm p} \simeq 0.5$).  
   The observed scatter may be taken as reflecting the different efficiencies of TDU to bring carbon into the envelopes of primary stars and/or attributed to the gravitational settling of accreted carbon in the envelope of secondary stars.  
   \citet{Weiss00} show that the heavy pollutants settle down to decrease their surface abundance by a factor up to $\sim 10$ in 10 Gyr, which may partly explain the spread of surface carbon abundances, observed from CH sub-dwarfs and also CEMP stars in Fig.~\ref{distCEMP+CH}.  
   The gravitational settling has little thing to do with the carbon enrichment of giants, however, since it is overtaken by the surface convection deepening along the red giant branch.  
   But note that for $\feoh \gtrsim -6$, the surface convection reaches deep enough to the shell where the hydrogen shell-burning passes, and hence, the accreted matter may affect the evolution of the secondary to the helium core flash \citep{Fujimoto95}.  

As for the separation (or the period), the observations lies well below the predictions based on Jeans' theorem.   
   Since the periods of the observed systems most fall in the range of eq.~(\ref{binary-shrinkP}) or shorter, they are thought to have suffered the orbital shrinkage during the mass transfer episode. 
   In order to see the effects of additional angular momentum loss, we figure out and plot the relations under the assumption of constant separation in the figure (denoted by broken lines), too;  they end with smaller final separation and larger accreted mass as compared to the relations based on Jeans' theorem.   
   The observed points mostly lie near the shortest ends of the separation necessary to contain the primary stars within the Roche robe at the beginning of TDU.  
   Furthermore, some of them including the two blue stragglers have the observed periods even shorter than the latter, which suggests that these systems have experienced a large shrinkage of separations during the common envelope phase.   

The top panel shows the results for CEMP stars, where the carbon abundance of the wind is taken to be $[{\rm C}/{\rm H}]_{\rm p} = 0$. 
   Since the radius of EMP primary at He-FDDM is smaller than the radus of Pop II primary at TDU (see Fig.~\ref{dependences}), the binary separation can be smaller for CEMP stars than CH stars. 
   CEMP stars of short period $P \lesssim 10$ yr must also have suffered the shrinkage of separations during the mass transfer episode. 
   If the shrinkage is taken into account, the relation calculated from the wind accretion scenario gives reasonable description to the observed trend of CEMP stars similarly to CH stars, discussed above. 
   It is worth noting that CEMP-nos star, CS22957-027 ($[\abra{C}{H}] = -7.1$ and $P = 3,125$ days) is also among the stars showing reasonable agreement with the theoretical prediction.  
   A dwarf HE0024-2523 of the shortest period, and possibly subgiant star CS29497-030 (names attached in the figure), have undergone RLO to accrete carbon-rich matter from the primary stars.  
   The eccentricities of these stars are nearly zero, which is also expected from RLO, though the detailed modeling of their evolution awaits future studies with the interaction between the stellar wind and the orbital motion properly taken into account.  
   The other dwarf G77-61 of the second shortest period may deserve comment because of rather weak carbon enrichment of $[\abra{C}{H}] \sim -1.4$ \citep{Plez05}. 
   It has the mass $\sim 0.3 \, M_{\odot}$ \citep{Dearborn86} and is likely to be wholly convective. 
   In addition, the cross section of the star for the accretion is a factor of $\sim 7$ smaller than that of a red-giant CEMP star [see Eq. (\ref{Bondi})], and hence, the accreted mass of carbon-rich material might be smaller by the same factor. 
   Even if we take into account these two factors that may lead to a relatively low value of [C/H], our model is difficult to reproduce the values of G77-61 with the carbon abundance $[\abra{C}{H}]_{\rm p} =0$. 
   Accordingly, this star requires both the shrinkage of orbital separation and the carbon abundance in the wind from the primary smaller than the solar abundance ($[{\rm C}/{\rm H}]_{\rm p} \simeq -1$).  

One of the factors that may affect these correlations is the difference in the CNO abundance in the wind between CEMP and CH stars, as discussed above.  
   The difference between CEMP dwarfs and CH dwarfs as large as $\sim 0.5$ dex is discernible also from the distribution of carbon abundances in Figs.~\ref{distCEMP+CH}.   
   This may reflect the difference in the efficiency of TDU due to the metallicity;  possible reason may be the increase in the hydrogen shell burning rate and the resultant expansion of envelope with increasing surface enrichment of carbon.  
   Another factor may be the eccentricity of the orbit. 
   The effect of the orbital eccentricity on the mass accretion rate can be incorporated as was done by \citet{Han95}, and a larger eccentricity tends to increase the mass accretion rate and thus increase the carbon abundance of the secondary. 

\section{Statistical Features of CEMP-s Stars }\label{statistics}

We have proved that our binary scenario gives a reasonable description of the observed characteristics for the particular CEMP stars with known binary periods in comparisons with CH stars. 
   Now we turn to the statistics of CEMP stars to demonstrate the validity of our binary scenario and to investigate insights into their properties, gained from our binary scenario, in the following two sections.  
   We first deal with CEMP-s stars of case II and II$^{\prime}$, which are well delineated as a subclass, both observationally and theoretically by the large enrichments not only of carbon and nitrogen but also of the s-process elements.  
   We then demonstrate that constraints can be drawn upon the initial mass function and the nature of EMP binaries with taking both CEMP-s and CEMP-nos stars into account.  

\subsection{Frequency in Comparison with CH Stars}\label{population}

\begin{figure}
\epsscale{1.0} \plotone{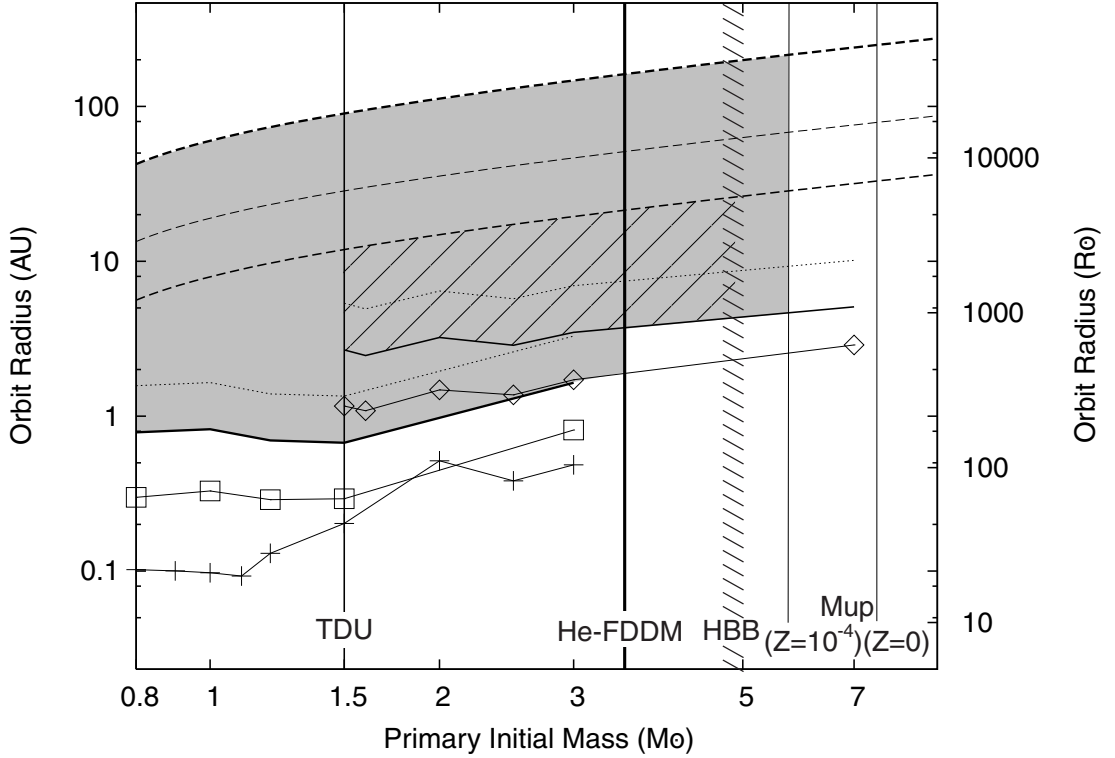}
\caption{Evolution to CEMP and CH stars in binary systems on the diagram of the initial mass of primary component and the initial separation (the mass of secondary component is set at $0.8 \msun$).  
   Symbols connected by thin solid lines represent the stellar radii of primary stars at the onset of He-FDDM and TDU, taken from the computations of $Z=0$ (crosses) and $\feoh = -3$ (boxes) models \citep[][Suda 2006 in preparation]{Suda04} and from the computations of more metal-rich populations \citep[diamonds][]{Iben75, Lattanzio86, Marigo02}. 
   Thick lines denote the orbital separations below which the primary fills its Roche lobe, leading to RLO before He-FDDM or TDU starts, and set the minimum initial separations for the surface carbon-enhancement. 
   Top three dashed lines denote the separations for which the envelope of secondary star is enriched with $[{\rm C}/{\rm H}] =-3$, $-2$, and $-1$ through the wind accretion; 
   here we assume that the wind is ejected from the primary with the solar carbon abundance and that the depth of surface convection of the secondary is $0.35 \msun$ in mass. 
   Four vertical lines denote, from left to right, the lower mass limit to TDU \citep{Lattanzio86}, the upper mass limit to He-FDDM \citep{Fujimoto00}, and the upper mass limits, $M_{\rm up}$, to TP-AGB evolution with C + O core for two metallicity of $Z = 10^{-4}$ and $Z=0$ \citep{Cassisi93}; 
   and vertical hatched-line indicates the lower mass limit to the hot bottom burning in the envelope \citep{Ventura01}.  
   Models of $\feoh = -3$ in shaded zones become CEMP via RLO or via wind accretion, and models of $\feoh = -1$ in hatched zones become CH stars via RLO or via wind accretion, respectively.  
   Thin dotted lines denote the upper separation limits to become C-rich via RLO. }
\label{dependences}
\end{figure}

On the basis of our scenario, we may estimate the fractions that CH and CEMP-s stars occupy among the Pop II and EMP stars, respectively. 
   The binary parameters that the secondary stars can evolve to CEMP-s and CH stars are depicted on the diagram of initial mass of primary stars and binary separation in Fig.~\ref{dependences}. 
   For CEMP-s stars, we may take the lower bound to the initial mass of primary stars to be $M_{1} (t_{\rm initial}) = 0.8 \msun$ and assume that the primary stars more massive than this mass have already ejected their carbon-rich envelope and evolved to white dwarfs. 
   The upper bound to the primary initial mass for CEMP-s stars is set at $M_{\rm He-FDDM} \simeq 3.5 \msun$ by the condition for He-FDDM to occur. 
   For CH stars, on the other hand, the lower and upper bounds are set by the lower mass limit to TDU at $M_{\rm TDU} \simeq 1.5 \msun$, and by the hot bottom burning (HBB), which converts carbon into nitrogen in the surface convection, at $M_{\rm HBB} \simeq 5 \msun$, respectively.   
   As for the initial binary separation, the first condition is that the primary star is allowed to develop carbon enhancement and sets the lower bounds, $A_{\rm He-FDDM}$ and $A_{\rm TDU}$, for CEMP-s and CH stars, respectively. 
   The second condition is to enrich enough the carbon abundance of the secondary star through the wind accretion, which imposes the upper bounds, $A_{M}(M_1, {\rm [C/H]}, M_{\rm scz})$.  
   The carbon enhancement necessary for CEMP and CH stars may be taken to be $[\abra{C}{H}] > -3$ and $-1$, respectively.  
   Consequently, CEMP-s stars occupy much larger parameter space than CH stars, as seen from Fig.~\ref{dependences}.  

In order to evaluate the relative frequencies of CEMP-s and CH stars numerically, we have to specify the initial mass function $\xi (M_1)$ for the primary stars and the distributions, $n(q)$ and $f(P)$, of binary systems with respect to the mass ratio $q \, (=M_2/M_1)$ and the orbital period $P$, respectively.  
   The fractions of CEMP-s and CH stars to EMP and Pop~II stars, $\phi_{\rm CEMP-s}$ and $\phi_{\rm CH}$, are then proportional to the following integrals:
\begin{eqnarray}\label{popcemp}
\phi_{\rm CEMP-s} & \propto & 
\int^{M_{\rm He-DDDM}}_{0.8 \msun} \xi(m) {n (M_2 / m) \over m} dm \nonumber \\
&  & \times \int^{A_{M}(m, [\abra{C}{H}],\, M_{\rm scz})}_{A_{\rm He-FDDM} (m)}  f(P) \frac{dP}{da}da, \\
\label{popch}
\phi_{\rm CH} & \propto &
\int^{M_{\rm HBB}}_{M_{\rm TDU}} \xi(m) {n (M_2 / m) \over m} dm \nonumber \\ 
&  & \times \int^{A_{M}(m, [\abra{C}{H}],\, M_{\rm scz})} _{A_{\rm TDU}(m)} f(P) \frac{dP}{da}da.
\end{eqnarray} 
   As for the initial mass function (IMF) for the primary stars of mass $> 0.8 \msun$, we may here assume Salpeter's power law, 
\begin{equation}
\xi (m)\propto m^{-2.35}.   
\end{equation}
   We may also assume a lognormal form as in the next section (see blow eq.\ (\ref{eq:lognormalIMF}). 
   For the binary periods, we may adopt the distribution which has been derived from the observations of local stars, 
\begin{equation}
f(P) \propto  \frac{1}{P} \exp \left( \frac{-\left(\log P-4.8\right)^2}{2 \times 2.3^2} \right), 
\end{equation}
\citep{Duquennoy91}, where $P$ is the period in units of days. 

On the contrary, as for the mass ratio distribution, no consensus has been achieved.
   For the near-by G-dwarf binary samples, it is shown that the distribution function can well be fitted by a Gaussian form, centered at $q = 0.23$ with a rather large dispersion of $\sim 0.42$, or could be even flat for $q < 0.23$.  
   It is also pointed out, however, that the same mass function as for the single stars in Millar and Scalo law is admissible for the secondary stars because of large uncertainty in the mass ratio below $q \lesssim 0.1$ \citep{Duquennoy91}.  
   Even with the recent studies it still remains an open question.  
   On one hand it is reported that the distribution have a small peak at $q \sim 0.8$ and rises as the mass ratio decreases to $q \simeq 0.2$ with a drop at $q <0.2$ for nearby spectroscopic binaries \citep{Goldberg03} or that it has a broad shallow peak between $q \sim 0.2$ and $\sim 0.7$ and a sharp peak at $q > 0.8$ for nearby F7-K stars \citep{Halbwachs03}.
   On the other hand \citet{Mazeh03} argue that the mass ratio distribution approximate constant over the range $q = 1.0-0.3$ and a constant distribution cannot be ruled out at lower $q$ for main-sequence binaries measured by infrared spectroscopy.  
   In particular, \citet{Goldberg03} find that the halo binaries have a flatter distribution than the disc binaries.  
   For such low mass binaries, the mass-ratio distribution is subject to large uncertainties in the range of mass ratio below $q \simeq 0.1$ since the main sequence does not exist for $M < 0.08 \msun$.  
   For massive star binaries with O-type and B-type primaries, in contrast, the mass function of the secondary stars is found to extend into the very small mass ratio $q < 0.1$ \citep{Abt83}.  
   In the present study, therefore, we may well assume a uniform distribution of mass ratio of binary systems for simplicity;
\begin{equation}
n(q)= {\rm const} \quad \hbox{ for } 0 < q \le 1.-
\end{equation} 
   The resultant mass functions of CEMP and CH stars depend only weakly on the secondary mass $M_2$ through the upper and lower boundary values of integral with respect to the separation.  
   This makes a sharp contrast with the case when the same single-star mass function as the primary stars is assumed for the secondary stars, as done by \citet{Lucatello05b}.  
   We may examine the validity of our assumption from the comparisons with the observed properties of CEMP and CH stars, as discussed later.  

The observed flux-limited samples of CEMP-s stars and CH stars are dominated by those which have masses at nearly upper end and are most luminous.  
   Putting $M_2 = 0.8 \msun$ and evaluating the double integrals with respect to the period and primary mass in eqs.~(\ref{popcemp}) and (\ref{popch}) numerically with $[\abra{C}{H}] = -3$ and $-1$ respectively for CEMP-s and CH stars, we find that:
\begin{equation}
\phi_{\rm CEMP-s}/\phi_{\rm CH}\simeq 14. 
\end{equation}
   This number represents the ratio between the frequencies of CEMP-s and CH stars among the stars of corresponding metallicity that are born in binaries.  
   Therefore, it gives reasonable account of the observed much larger frequency of CEMP stars among EMP stars as compared with the frequency of CH stars among stars of Pop II stars, unless the fraction of EMP stars born in binaries is significantly smaller than those of the more metal-rich populations.  
   In evaluating the ratio, we have applied the Salpeter's IMF to EMP stars as well as to CH stars, although there is no guarantee that both obey the same initial mass function.  
   Nevertheless, this conclusion will be little affected since the large ratio arises essentially from the difference in the parameter space that EMP and Pop II stars can develop the carbon-enhancement, as seen from Fig.~\ref{dependences}.  
   We will discuss the influence of different IMFs in the next section.  

\subsection{Period Distributions in Comparison with CH Stars }\label{periodD}

\begin{figure}
\epsscale{1.0} \plotone{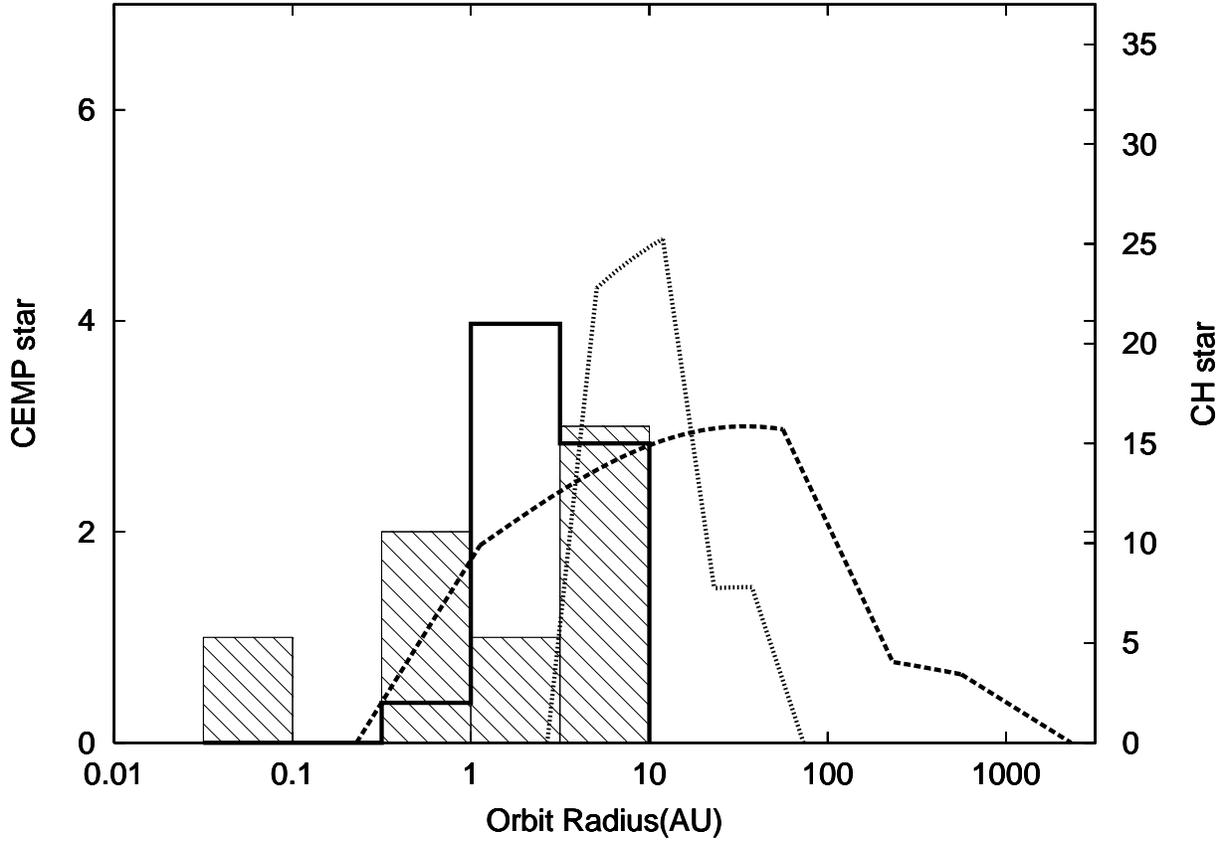}
\caption{Distributions of orbital separations observed from CEMP-s stars (hatched columns) and from CH stars (columns enclosed by thick solid lines). 
   Dashed and dotted curves depict the relative distributions of initial orbital separations, predicted from the binary scenario for CEMP-s stars and for CH stars, respectively: 
   a shoulder on the right of each curve comes from the contribution of dwarfs. 
} \label{OrbitD}
\end{figure}

Since CEMP-s stars and CH stars stem from the binary systems of different parameters in our binary scenario, they should differ in the distribution of the orbital separation (or the period).  
   The relative frequencies of the initial orbital separation (or the initial orbital period) may also be predicted from the integration of eqs.~(\ref{popcemp}) and (\ref{popch}) with respect to the primary mass for CEMP stars and CH stars, respectively.   
   The derived relative frequencies are plotted in Fig.~\ref{OrbitD} and compared with the observations.  
   Here, we take into account both giants and dwarfs and a shoulder on the right-hand side of each curve comes from the contribution of dwarfs.  

For CH stars, our binary scenario predicts the initial separations (or periods) restricted to rather narrow range  between $3 \sim 30$ AU (${\rm several} \sim 100$ years).  
   The observed periods also fall in a narrow range of separations (or orbital periods) in agreement with the prediction, but are shifted toward smaller separations (or shorter periods) by several factors on average.  
   This shrinkage of binary systems is consistent with the argument in the preceding section, based on the eqs.~(\ref{binary-shrinkS}), which anticipates that the wind may carry away larger amount of angular momentum than specific to the orbital motion of primary star in interaction with the secondary motion.  

In contrast, CEMP-s stars are predicted to stem from the binaries of much larger range of initial separations (or initial periods);  
   for those with $[\abra{C}{H}] > -3$, the initial parameters range from 0.3 AU (less than a year) to more than 1000 AU ($\sim 10000$ years) with the central values near $\sim 30$ AU ($\sim 100$ years). 
   All the observed orbital periods of CEMP-s stars are less than $\sim 10$ yrs because of limited span of monitor time to date, and fall within the shorter half of initial period range, predicted from our scenario.  
    For such shorter periods, the binary systems must have suffered from shrinkage during the mass loss event from the primaries. 
    The observations are consistent with this expectation, where the distribution of observed separations (or periods) is extended toward shorter separations (or periods) than predicted. 
    In addition, we expect from Figs.~\ref{OrbitD} and \ref{dependences} that more than a half of CEMP-s giants, and still larger portion of CEMP-s dwarfs, have wider separations (or longer periods) than observed to date.
  In this parameter range, Jeans' theorem tends to hold and the binary separations (or periods) further increase.
  So, in the extreme end, it is difficult to detect the orbital motion of binary spectroscopically as a periodic variation of absorption lines. 
   
In summary, the binary scenario is compatible not only with the observed period distribution of CEMP-s stars but also with the lack of detected variations in the radial motion for a half of them to date. 
   These results are indifferent to the assumptions on the initial mass function and on the distribution of mass ratio.   
   They will not be basically influenced by the assumption on the period distribution, $f(P)$, either, unless it varies greatly in the range of separation, delineated as the parameter space for CEMP-s stars in Fig.~\ref{dependences}.  

\subsection{Frequencies of Dwarfs and Giants}

CEMP stars make a sharp contrast with CH stars in their relative distributions with respect to the evolutionary stages, too. 
   Among CEMP-s stars, giants overweigh dwarfs and subgiants in number (14 giants among 21 stars in our sample), while subgiant CH stars outnumber CH (giant) stars (18 CH giant stars and 25 subgiant CH stars in our sample).  
   From the binary scenario, this may be explicable in terms of the effect of the pristine metal abundances in the dilution of carbon-enrichment when a star evolves to be a giant. 
   For such large pristine metallicity as CH stars, the mixing of accreted carbon-rich matter into the extended surface convection of giants is ready to convert the carbon-rich dwarf stars to the giant stars exhibiting no enhancement of carbon, i.e., $\abra{C}{O} < 1$; 
   we see in Fig.~\ref{FeC} that CH giants are predominant at lower metallicity; 
   if Ba stars of $\abra{C}{O}<1$ are included, giants may exceed dwarfs in number (see Fig.~\ref{distCEMP+CH}) although we should take into account the differences in the metallicity dependence \citep[e.g., see][]{Smith86}. 
   For CEMP stars, on the contrary, the carbon enrichment may be discernible even if dreaded-up carbon is diluted in the surface convection of giants of mass larger by a factor of $\sim 100$ than the dwarfs.  

In the binary scenario, the mass distribution of CEMP-s stars is expressed from eq.~(\ref{popcemp}) as; 
\begin{eqnarray}
\xi_{\rm CEMP-s} (M_2)
& = & \int^{M_{\rm He-FDDM}}_{0.8\,M_\odot} {d m} \xi (m) \frac{n(M_2/m)}{m}  \nonumber \\
& & \times \int^{A_{M}(m,-3,\, M_{\rm scz})}_{A_{\rm He-FDDM}(m)} f(P) \frac{dP}{da}da.
\end{eqnarray} 
   We may estimate the ratio between giants and dwarfs from our scenario by using the results of evolutionary study of a star of mass of $0.8 \ M_\odot$ for the metallicity $\feoh = -3$ (Suda 2006 in preparation). 
   It takes $ t_{\rm f} = 15.9$ Gyr for this star to evolve from the zero-age main sequence to a white dwarf, including 1.3 Gyr $(=\Delta t_{\rm SG} = t_{\rm g} - t_{\rm TO})$ spent as subgiant after the turn-off (at $t = t_{\rm TO}$) to the stage when the surface convective zone becomes deep (at $t = t_{\rm g}$), and then, 0.45 Gyr ($=\Delta t_G  = t_{\rm f} - t_{\rm g}) $ before it becomes a white dwarf at $t=t_{\rm f}$. 
   The mass ranges, $\Delta m_{\rm SG}$ and $\Delta m_{\rm G}$, of subgiant and giant stars that were born in early days ($\sim t_{\rm f}$ years ago) can be estimated from the mass-luminosity relation for dwarf stars and the time spans during which the star resides as a subgiant and on the RGB and horizontal branch, respectively. 
   The mass-luminosity relation for dwarf stars is assumed to follow a power law $L_{\rm dwarf}(m) \propto m^\alpha \, (\alpha = 3.5)$. 
   From the relation 
\begin{equation}
{\Delta t_G}/{t_ {\rm f}} = (\alpha-1) \left({\Delta m}/{0.8M_\odot }\right),
\end{equation} 
   we obtain $\Delta m_{\rm G} = 0.01\,M_\odot$ and $\Delta m_{\rm S G}=0.03 \, M_\odot$. 

Since the observed EMP stars constitute flux-limited samples, we should also take into account the differences in the luminosity among them and their spatial distributions. 
   We denote, by $N_{\rm s}(L)$, the surface number density (per unit steradian) of stars with the luminosity $L$ that are accessible with a given limiting flux $f_{\rm limit}$. 
   As for the special distribution, we may well assume that the EMP stars obey the density distribution, $\rho(r)$, of stars in the Galactic halo, approximated by 
\begin{equation}
\rho(r) \propto r^{-3}
\end{equation} 
   as a function of radial distance $r$ from the Galactic center \citep[e.g. see][and references therein] {Majewski93}. 
    Since the surveys have been performed exclusively in the fields of high galactic latitudes, we may well consider the direction perpendicular to the Galactic disk, which leads to;   
\begin{eqnarray}
N_{\rm s}(L) & \equiv & \int^{\sqrt{{L}/{4\pi f_{\rm limit}}}}_{0} \rho (r) z^2 d z \nonumber \\ 
             & \propto & \int^{\sqrt{{L}/{4\pi f_{\rm limit}}}}_{0} \frac{z^2}{\left(8^2 + z^2\right)^{3/2}} d z, 
\end{eqnarray}
   where the distance from the Galactic center to the sun is taken to be 8 kpc and $z$ denotes the height above Galactic disc in unit of kpc. 
   This factor becomes important as the stars evolve to be luminous along the giant branch.  
   For the dwarfs with low luminosity, on the other hand, the spatial variations in the density of EMP stars are negligible and we simply have $N_s (L) \propto L^{3/2}$.  

The size of surface convection, $M_{\rm SCZ}$, differs by a factor of $\sim 100$ (see eq.~[\ref{am}]) between dwarfs and giants, and hence, the upper limit, $A_M (M_1, [\abra{C}{H}],\, M_{\rm scz})$, to the binary separation of carbon enhancement for dwarfs and subgiants is 10 times larger than that for giants.  
   Thus, the number ratio of CEMP-s giants to CEMP-s dwarfs and subgiants will be given by $\psi_{\rm s, giant} / (\psi_{\rm s, dwarf} + \psi_{\rm s, subgiant} )$, each of which is defined as; 
\begin{eqnarray}
\psi_{\rm s, dwarf} & = & \int^{0.8}_{0.08} dM_2 N_{\rm s}(L(M_2)) \nonumber \\ 
   & & \times \int^{M_{\rm He-FDDM}}_{0.8} dM_1  \xi (M_1) \frac{n(M_2/M_1) }{M_1}  \nonumber \\  
   & & \times \int^{A_{M}(M_1,-3,\, M_{\rm scz,D})}_{A_{\rm 
He-FDDM}}f(P)\frac{dP}{da}d a, \label{eq:num-D}\\
\psi_{\rm s, subgiant} & = &  \int^{0.8 + \Delta m_{SG}}_{0.8} dM_2 \frac{\int^{t_{\rm g}}_{t_{\rm sg}} N_{\rm s}(L(t)) d t }{t_{\rm g}-t_{\rm sg}}  \nonumber \\ 
   & & \times \int^{M_{\rm He-FDDM}}_{0.8} dM_1 \xi (M_1) \frac{n(0.8/M_1) }{M_1} \nonumber \\ 
   & & \times \int^{A_{M}(M_1,-3 ,\, M_{\rm scz, D})} _{A_{\rm He-FDDM}}f(P)\frac{dP}{da}d a. \label{eq:num-SG}\\
\psi_{\rm s, giant} & = & \int^{0.8 + \Delta m_{SG}+ \Delta m_G}_{0.8+ \Delta m_{SG}}  dM_2 \frac{\int^{t_{\rm f}}_{t_{\rm g}} N_{\rm s}(L(t)) d t }{t_{\rm f}-t_{\rm g}} \nonumber \\ 
   & & \times \int^{M_{\rm He-FDDM}}_{0.8} dM_1  \xi (M_1) \frac{n(0.8/M_1) }{M_1} \nonumber \\ 
   & & \times \int^{A_{M}(M_1,-3 ,\, M_{\rm scz, G})}_{A_{\rm He-FDDM}}f(P)\frac{dP}{da}d a. \label{eq:num-RG}
\end{eqnarray} 
   By adopting the limiting magnitude of 15 mag, we find the resultant proportions of RGB+HB stars, subgiants, and dwarfs at 63 \%, 10 \%, and 27 \%, respectively. 
   In our samples, there are 13 giants, 4 subgiants, and 4 dwarfs (62 \%, 19 \%, and 19 \%, respectively) among CEMP-s stars if we define dwarfs as stars with the surface gravity, $\log g \geq 4.2$, and subgiants as  $ 4.2 > \log g \geq 3.5$, and giants (including HB stars) as $\log g < 3.5$, respectively.  The above estimate based on the binary scenario gives a reasonable agreement with the current observations.   

\section{Initial Mass Function and Binarity of EMP Stars}\label{IMFsec}

In the preceding section, we have demonstrated that the statistical features of CEMP-s stars as a subclass are understandable in terms of our binary scenario. 
   In this section, we take up the problems of the observed frequencies of CEMP-s and CEMP-nos stars among EMP stars and of the total number of EMP stars in an attempt to inquire into the initial mass function (IMF) of EMP stars and to give an insight into their nature.  

From our scenario, we can estimate the proportions of CEMP-s and CEMP-nos stars to EMP binaries. 
   The number, $\psi_{\rm binary}$, of all the EMP stars born in binary systems in a flux-limited sample may be evaluated in the same way as we have derived the CEMP-s population in the preceding section.  We may divide the EMP binaries into three categories; 
   in addition to \hfill\break
   (1) white-dwarf binary systems, so far discussed, in which the primary stars in the mass range of $0.8 \msun < M_1 \le M_{\rm up}$ have already become white dwarfs and the secondary stars still undergo nuclear fusion, we define two more of  \hfill\break
   (2) low-mass binary systems in which both of the members are in the mass range of $ \le 0.8 \msun$ and still alive as nuclear burning stars, and  \hfill\break 
   (3) supernova binary systems in which the primary stars in the mass range of $M_1 > M_{\rm up}$ have already exploded as supernovae and the secondary stars still remains on nuclear burning stages. \hfill\break 
   Then the total EMP stars currently observed are given by the sum of these three categories as; 
\begin{equation}
\psi_{\rm binary} =  \psi_{\rm WDB } + \psi_{\rm LMB} + \psi_{\rm SN}, 
\label{eq:EMP}
\end{equation}
  where 
\begin{eqnarray}
\psi_{\rm WDB } & = & \int^{0.8 \, M_\odot}_{ 0.08 \, M_\odot}dM_2 N_s (L[M_2]) \int_{0.8\,M_\odot}^{M_{\rm up}} dM_1  \xi (M_1) \frac{n(q)}{M_1} \nonumber \\  
   & & \times \int^{A_{\rm cut}}_{A_{\rm min, WD} (M_1, M_2)} f(P) \frac{dP}{da}da \\
\psi_{\rm LMB} & = &  \int^{0.8 \, M_\odot}_{0.08 \, M_\odot} d M_2\int^{0.8 \, M_\odot}_{ M_2 } dM_1  N_s (L[M_1]) \xi (M_1) \frac{n(q)}{M_1} \nonumber \\  
   & & \times \int^{A_{\rm cut}}_{A_{\rm min, MS} (M_1, M_2)} f(P) \frac{d P}{d a} d a, \\
\psi_{\rm SN } & = & \int^{0.8 \, M_\odot}_{ 0.08 \, M_\odot}dM_2 N_s (L[M_2]) \int^{\infty} _{M_{\rm up}} dM_1  \xi (M_1) \frac{n(q)}{M_1} \nonumber \\  
   & & \times \int^{A_{\rm cut} }_{A_{\rm min, SN} (M_1, M_2)} f(P) \frac{dP}{da}da .
\end{eqnarray} 
   Here ${A_{\rm min, WD} (M_1, M_2)}$, ${A_{\rm min, MS} (M_1, M_2)}$ and ${A_{\rm min, SN} (M_1, M_2)}$ denotes the minimum separations that can contain two stars of AGB and main sequence, both of main sequence stars, and of supernova progenitor and main sequence star, respectively:  
   $A_{\rm cut}$ denotes the upper bound of binary separation, and is taken to be consistent with $P = 10^{10}$ days given by \citet{Duquennoy91}. 

The evaluation of these numbers requires the initial mass function of EMP stars in the whole mass range, and we may well assume a lognormal form of IMF with the medium mass, $M_{\rm md}$, and the variance, $\Delta_M$, as parameters, i.e., 
\begin{equation}
\xi (m)\propto \frac{1}{m} \exp \left( -\frac{(\log m-\log M_{\rm md})^2}{2\times \Delta _M ^2} \right). 
\label{eq:lognormalIMF}
\end{equation}
   For the Galactic spheroid, the IMF is shown to be well approximated by a lognormal form with the parameters of $M_{\rm md} = 0.22 \msun$ and $\Delta _M = 0.33$ for low mass stars ($m<1\msun$) and Salpeter form for more massive stars \citep[see e.g., review by][]{Chabrier03}. 
   EMP stars may not necessarily follow the same IMF as the Galactic spheroid components of larger metallicity. 
   Rather we here treat these two parameters as free and discuss the constraints imposed on them from the comparisons with the current observations. 
   As for the binary mass ratio and period, we assume the same distributions as adopted in the preceding section.  

\subsection{Frequency of CEMP-s Stars}

\begin{figure}
\epsscale{1.0} \plotone{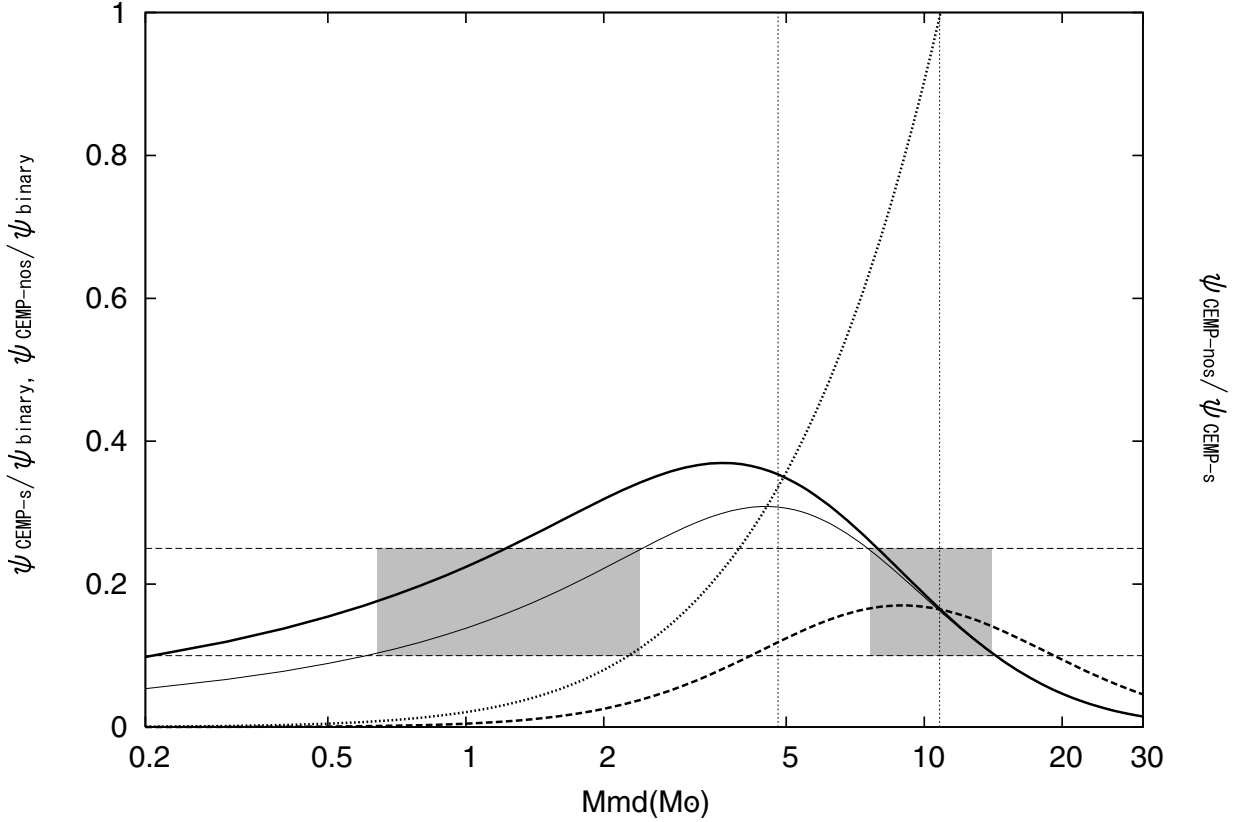}
\caption{
Fractions of CEMP-s and CEMP-nos stars among EMP stars as a function of the medium mass $M_{\rm md}$, a parameter of the initial mass function in a log-normal form of eq.~(\ref{eq:lognormalIMF}) with the other parameter fixed at $\Delta _M = 0.33$. 
   Thick solid and broken lines plot the proportions of CEMP-s and CEMP-nos stars to EMP stars born in binaries, $\psi_{\rm CEMP-s} / \psi_{\rm binary}$ and $\psi_{\rm CEMP-nos} / \psi_{\rm binary}$, predicted from the binary scenario, respectively:  
   Thin solid lines depicts the CEMP-s proportion when we take into account the contribution from low-mass stars born as single stars at the equal number with the binaries.  
   Thick dotted line denotes the number ratio of CEMP-nos to CEMP-s stars, $\psi_{\rm CEMP-nos} / \psi_{\rm CEMP-s}$.   
   The observed proportion of CEMP-s stars ($\simeq 10 \sim 25 \%$) is bounded by two horizontal thin broken lines and two shaded areas indicate the parameter ranges of $M_{\rm md}$ compatible with the observations.  
   Two vertical thin dotted lines demarcate the observed bounds of the ratio between CENP-nos and CEMP-s stars ($\simeq 1/3 \sim 1$).  
}
\label{s/nos}
\end{figure}

The population of CEMP-s stars, $\psi_{\rm CEMP-s}$, predicted from our binary scenario, is given by the sum of CEMP-s dwarfs, subgiants and giants (including HB stars) in eqs.~(\ref{eq:num-D})-(\ref{eq:num-RG}); 
\begin{equation}
\psi_{\rm CEMP-s} = \psi_{\rm s, dwarf} + \psi_{\rm s, subgiant} + \psi_{\rm s, giant}. 
\label{eq:CEMP-s}
\end{equation} 
   We may integrate eqs.~(\ref{eq:CEMP-s}) and (\ref{eq:EMP}) by use of IMF in eq.~(\ref{eq:lognormalIMF}).  
   Figure~\ref{s/nos} plots the proportion of CEMP-s stars to EMP binaries, $\psi_{\rm CEMP-s} / \psi_{\rm binary}$, as a function of medium mass, $M_{\rm md}$, with a fixed dispersion of $\Delta _M = 0.33$, assuming simple lognormal IMFs.  
   If the same initial mass function as the Galactic Spheroid (peaked at $0.22 \msun$) is applied, we find 
\begin{eqnarray}
& & \psi_{\rm CEMP-s} / \psi_{\rm WDB} \simeq 0.58: \nonumber \\
& & \psi_{\rm CEMP-s} / \psi_{\rm binary} \simeq 0.14. \nonumber
\end{eqnarray}
   The former ratio indicates that more than a half of white-dwarf binaries have produced CEMP-s stars, which is attributable to the very large coverage in the parameter space of separations and mass ratios, as stated above. 
   Nevertheless, their proportion to the total EMP stars decreases by a factors of $\sim 4.1$ owing mainly to the contribution from the low-mass binaries. 
   If we take into account the contribution of the stars born as single, this fraction is too small when compared with the observed very high frequencies of CEMP-s stars (see below).  
   This may be indicative of more massive IMF for EMP stars.  
   As $M_{\rm md}$ increases, the frequency of low-mass binaries diminishes and the CEMP-s proportion increases to attain at a maximum of $\psi_{\rm CEMP-s} / \psi _{\rm binary} \simeq 0.37$ near $M_{\rm md} \simeq M_{\rm He-FDDM}$. 
  For a still larger $M_{\rm md}$, it turns to decrease as the contributions augment from white-dwarf binaries with the massive primaries of $M > M_{\rm He-FDDM}$ and then from supernova binaries of $M > M_{\rm up}$.  
   In this figure, we also plot the CEMP-s proportion if we assume that the same number of EMP stars are born as single.  
   The contribution of single-born EMP stars may be significant for the IMF of small $M_{\rm md}$, but decreases rapidly to be negligible for the IMF of large $M_{\rm md} > M_{\rm He-FDDM}$.  

From the HK and HES observations, it is reported that the proportion of CEMP stars of $[\abra{C}{Fe}] >1$ amounts to $25\%$ for in the metallicity range of $\feoh < -2.5$ \citep[][see also Lucatello et al.~2005a,b]{Beers99,Rossi99,Christlieb03}.  
   They include the contribution from CEMP-nos stars, which accounts for $\sim 25\%$ of CEMP stars \citep{Ryan05,Aoki06}. 
   If allowance is made for the contribution of CEMP-nos, then, the CEMP-s proportion may be smaller, while larger if we include such CEMP-s stars as having smaller carbon enrichment of $1> [\abra{C}{Fe}] > 0.5$.   
   Recently, \citet{Cohen05} find the frequency of CEMP stars at $14.4\pm 4 \%$ for $\feoh < -2$ and $[\abra{C}{Fe}] \ge 1.0$ in HES sample, by correcting errors in abundance analyses of CEMP stars by means of the follow-up spectroscopy.  
   This gives the CEMP-s fraction at $\sim 10$\% with the possible contribution from CEMP-nos subtracted.
   However, their definition of CEMP stars is different from ours and may underestimate the CEMP-s fraction since the efficiency of producing the carbon enrichment greatly decreases for larger metallicity of $\feoh > -2.5$, as discussed above. 

Accordingly, we may well adopt the observed proportion of CEMP-s stars among the EMP stars at $10 \sim 25$\%.  
   As seen from Fig.~\ref{s/nos}, there are two possible solutions with the separate ranges of $M_{\rm md}$ that predict the proportion of CEMP-s stars compatible with the observations;  one is the low-mass IMF with $M_{\rm md} \simeq 0.6 \sim 2.5 \msun$ and the other the high-mass IMF with $M_{\rm md} \simeq 7.5 \sim 14 \msun$.  
   The two solutions are separated by the overproduction of CEMP-s stars from the white dwarf binaries because of large coverage in the parameter space, as seen above from Fig.~\ref{dependences}.  
   Although the both IMFs produce the similar proportions of CEMP-s stars, they predict quite different constituents for other EMP stars.  
   In the case of the low-mass IMF, EMP stars other than CEMP stars are constituted mainly of low-mass stars born as single and of the members of low-mass binaries.  
   In the case of the high-mass IMF, the extant EMP stars were most born as the low-mass member of binaries and more than a half of them stem from the binaries of massive primary, which have undergone supernova explosions.  

Our result differs from that of \citet{Lucatello05b}, who find only the low-mass solution with $M_{\rm md} \simeq 0.79 \msun$, which arises from the different assumption on the distribution of mass fraction $n (q)$; 
  we adopt a flat distribution of mass fraction while they assume that the primary and secondary stars both obey the same initial mass function of single stars.   
   As stated above, either of these two assumptions is not ruled out according to the current researches, and yet, they entail distinct consequences on the mass spectra of secondary stars.  
   We have a nearly flat mass function for CEMP and also for CH stars. 
   On the other hand, if we apply their assumption to CH stars, it entails the predominance of low mass stars down to $\sim 0.2 \msun$ \citep{Chabrier03}.  
   From the observation of CH stars, however, the contrary is reported that subgiant CH stars are found only in a narrow spectral-type range around G0 but not among dwarfs of late G and K-types \citep{Luck82}. 
   This lack of low-mass main-sequence CH stars favors our assumption of a flat distribution of mass ratio rather than the same single-star mass-function of primary and secondary stars.  
   \citet{Luck91} have argued the thick accreted layers as the cause, but it seems difficult to prevent the formation of low mass CH stars through the wind accretion, as seen in \S~\ref{period}.  
   These assumptions are distinguished also in the production of CEMP-nos, as discussed below. 

\subsection{Relative Frequencies of CEMP-s and CEMP-nos Stars}\label{IMF-subsc}

In our scenario, both CEMP-s and CEMP-nos arise from the white dwarf binaries, for which the existence and absence of s-process element enhancement are separated by the mass of primary components at $M_1 = M_{\rm He-FDDM}$.
   Then, the populations of CEMP-nos stars can be estimated in the similar way to the populations of CEMP-s stars as; 
\begin{equation}
\psi_{\rm CEMP-nos} = \psi_{\rm nos, dwarf} + \psi_{\rm nos, giant} + \psi_{\rm nos, subgiant}, 
\label{eq:CEMP-nos}
\end{equation} 
   where the each term in the right-hand side member can be computed by integrating the corresponding equations of (\ref{eq:num-D})-(\ref{eq:num-RG}) for the mass range of $M_1$ between $M_{\rm He-FDDM}$ and $ M_{\rm up}$ and by replacing the lower bound of separation by $A_{\rm TDU}$.  
   Because of difference in the primary mass, the ratio between $\psi_{\rm CEMP-s}$ and $\psi_{\rm CEMP-nos}$ depends on the assumed initial mass function.  
   If we assume the Salpeter IMF for $M > 0.8 \msun$, it reduces to 
\[\psi_{\rm CEMP-nos} / \psi_{\rm CEMP-s}\simeq 1/50, \]
   which predicts a negligible fraction of CEMP-nos stars as compared with CEMP-s stars.  
   The distribution, $n(q)$, of mass fraction is unlikely to differ by an order of magnitude in the mass range of concern here and the distribution, $f(P)$, of orbital period is also expected not to vary greatly with the primary mass.  
   Accordingly, larger population of CEMP-nos stars is possible only for the initial mass function shifting toward higher mass. 

In Fig.~\ref{s/nos}, we plot the proportion of CEMP-nos stars to the EMP binaries and the ratio between the CEMP-nos and CEMP-s stars, as a function of $M_{\rm md}$.  
   As $M_{\rm md}$ increases and the IMF shifts to be more massive, the CEMP-nos proportion first increases, and hits a maximum of $\psi_{\rm CEMP-nos} / \psi_{\rm binary} \simeq 0.17$ near $M_{\rm md} \simeq M_{\rm up}$. 
   The maximum fraction is smaller than that obtained for CEMP-s stars above because of larger mass ratio between the primary and secondary stars. 
   For a still larger $M_{\rm md}$, the CEMP-nos fraction turns to decrease and the supernova binaries outnumber the other type binaries.  
   The ratio, $\psi_{\rm CEMP-nos}/ \psi_{\rm CEMP-s}$, is a monotone increase function of $M_{\rm md}$;  it starts from a very small values for the Salpeter IMF, the increase is accelerated for $ M_{\rm md} > M_{\rm He-FDDM}$ because of the decrease in the CEMP-s proportion, and eventually, the CEMP-nos proportion exceeds the CEMP-s proportion for $M_{\rm md} > 11 \msun$. 

The ratio between CEMP-nos and CEMP-s stars is reported to be $\simeq 1/3$ if limited to $[\abra{C}{Fe}] >1$ \citep[Aoki et al.~2006 in preparation;][]{Tsangarides04}.  
    In our compilation of $[\abra{C}{Fe}] > 0.5$, the ratio between CEMP-nos to CEMP-s stars is $13 / 21 = 0.62$, with two stars rich in r-process elements included.  
    In the above estimates, the upper mass bound, $M_{\rm up}$, of primary stars for the binaries producing CEMP-nos stars is set equal to the lower mass-limit at which carbon ignite under the non-degenerate condition.  
     Because of uncertainties in modeling of hot bottom burning (e.g., due to dependence on the mixing length), however, the boundary may not be clearly delineated from those yielding EMP stars that have low or moderate carbon enhancement but show large nitrogen enrichment.  
     In particular, most of the ``mixed'' stars, discussed by \citet{Spite05} are left out from our compilation since we define CEMP stars as $[\abra{C}{Fe}] > 0.5$.  
   These nitrogen-rich stars could be considered as the counterpart of CEMP-nos stars with more massive primary, and hence, processed deeply by the hot bottom burning.
   If ``mixed'' stars are included, the number of CEMP-nos stars amounts to 27 and is comparable with, or even exceeds, the number of CEMP-s stars in our list.   
   Accordingly the current observations suggest the ratio of CENP-nos to CEMP-s stars in the range of $\psi_{\rm CEMP-nos}/ \psi_{\rm CEMP-s} \gtrsim1/3$ and we may well take an upper bound at $\sim 1$.  
   The ratios in this range agree well with those predicted from the high-mass IMF with $M_{\rm md} \simeq 10 \msun$, one of the two solutions derived above from the CEMP-s proportion.
   The low-mass IMF is excluded on the basis of current observations since it can predict too small proportion of CEMP-nos stars ($\psi_{\rm CEMP-nos}/ \psi_{\rm CEMP-s} < 0.05$).  

\subsection{Nature of EMP Stars in Galactic Halo}\label{NatureEMP}  

\begin{figure}
\epsscale{1.0} \plotone{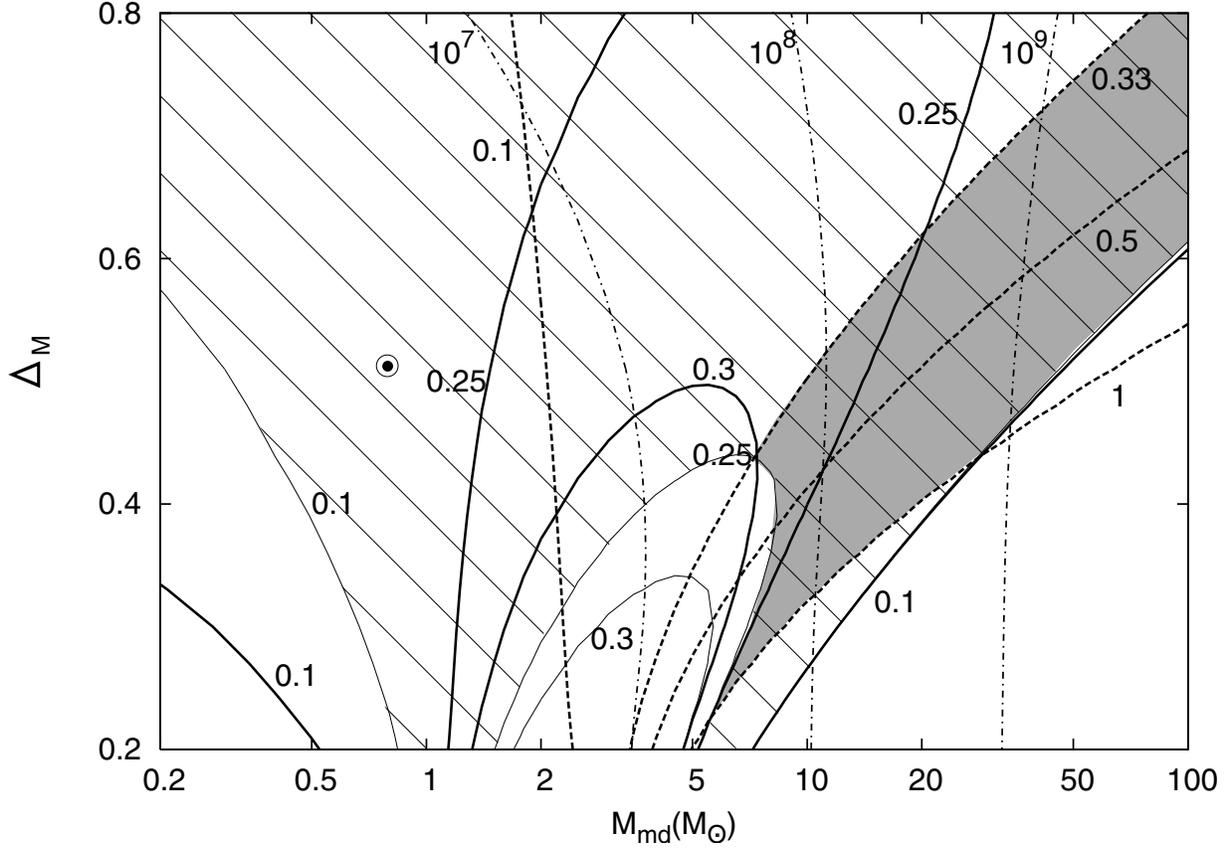}
\caption{
The initial mass function of EMP stars on the parameter diagram of the medium mass, $M_{\rm md}$, and the dispersion, $\Delta _M$, in a log-normal form of eq.~(\ref{eq:lognormalIMF}).  
   Thick solid lines depict the loci of constant proportions of CEMP-s stars to EMP binaries, $\psi_{\rm CEMP-s} / \psi_{\rm binary} \ ( = 0.1$, 0.25, and 0.3): 
   thin solid lines denote the loci of constant CEMP-s fractions when the EMP stars born as single stars at equal numbers to the binaries are taken into account, and hatched area delineates the parameter space of IMFs that give the fraction of CEMP-s stars consistent with the observations.   
   Broken lines denote the loci of constant number ratios between CEMP-nos and CEMP-s stars, $\psi_{\rm CEMP-nos} / \psi_{\rm CEMP-s} \ (= 1, 0.5, 0.33)$, and shaded area defines the parameter space that entails the IMFs compatible with the observations both of CEMP-s and CEMP-nos stars. 
   Thin dash-dotted lines denote the loci of constant total mass, $M_{\feoh < -2.5}$, of mother stellar populations $( = 10^7, \ 10^8$ and $10^9 \msun$) necessary to yield such the number of EMP stars, now constituting the Galactic halo, as estimated from the recent HES survey under the different assumptions of IMF.  Double circle denotes the IMF derived by \citet{Lucatello05b}}
\label{mcsigma}
\end{figure}

We have shown that the IMF for EMP binaries can be constrained from the observed characteristics of CEMP-s and CEMP-nos stars. 
   The constraints thus derived are summarized in Figure~\ref{mcsigma} on the diagram of the medium mass $M_{\rm md}$ and the standard deviation $\Delta_M$, the parameters of initial mass function in the lognormal form of eq.~(\ref{eq:lognormalIMF}). 
   Solid lines denote the loci of the IMFs that yield CEMP-s stars at constant fractions, and broken lines the loci of IMFs that produce CEMP-s and CEMP-nos at constant ratios.  
   Since the production rate of CEMP-s stars from the white dwarf binaries is very high ($\sim 60\%$), the parameter space near $M_{\rm md} \simeq M_{\rm He-FDDM}$ is excluded by the overproduction of CEMP-s as compared with the observation for the dispersion of $\Delta_M \lesssim 0.4$.  
    For smaller dispersion, therefore, the allowed parameter space for CEMP-s stars is separated into two parameter spaces with low and high mass regimes, converging toward $M_{\rm md} \simeq 1 $ and $6 \msun$ respectively.  
   As $\Delta_M$ increases, the wider range of $M_{\rm md}$ comes to be compatible with the observation of CEMP-s stars, and two regimes are connected for such broad IMF as $\Delta_M > 0.4$.  
   On the other hand, the formation of CEMP-nos stars at the observed ratios to CEMP-s stars restricts the IMFs to those with the large $M_{\rm md} \ (> M_{\rm He-FDDM}$); 
   the allowed range of $M_{\rm md}$ increases as the IMF grows broader and gives birth to larger fraction of low mass stars.  

\begin{figure}
\epsscale{1.0} \plotone{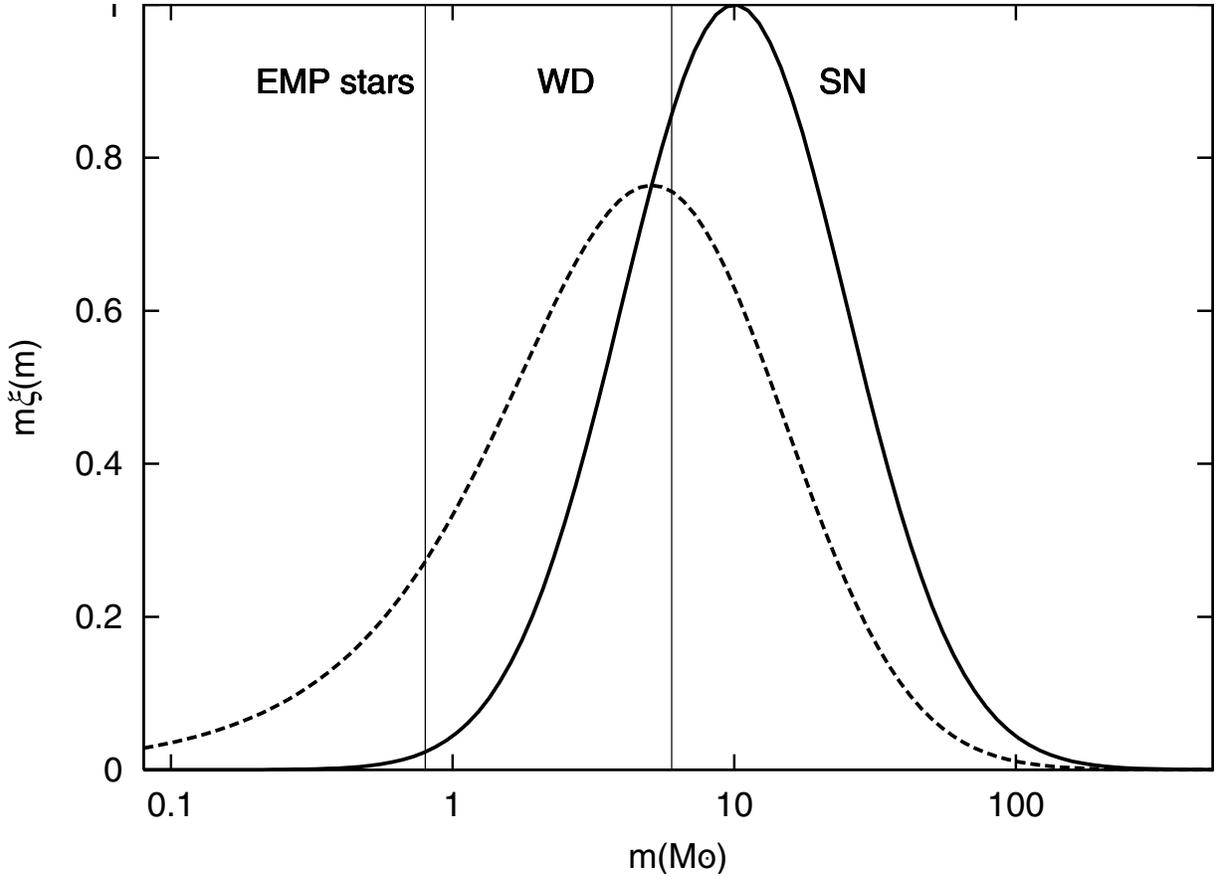}
\caption{
An illustration of relative distributions, $m \xi (m)$ of stellar masses for the derived initial mass function of EMP stars with $M_{\rm md} = 10 \msun$ and $\Delta_M = 0.4$;  solid and broken curves denote the mass distributions of primary components and secondary components, respectively.  }
\label{IMF}
\end{figure}

In conclusion, only compatible with the observations both of CEMP-s and CENP-nos stars are the IMFs with $M_{\rm md}$ greater than $\sim 6 \msun$ and increasing with $\Delta_M$, as shaded in Fig.~\ref{mcsigma}.  
   One of the important consequences is that the currently observed EMP stars have to stem exclusively from the binary systems;   
   Figure~\ref{IMF} exemplifies the mass distributions of primary and secondary stars for a typical initial mass function derived here. 
   We see that the secondary components well extend into low mass regime below $0.8 \msun$ but the primary components have a negligible fraction of low mass stars.
   We may not expect a significant fraction of low mass stars born as single unless the initial mass function for single stars may greatly differs from that of primary stars.  
   Another remarkable consequence is that a significant fraction of currently observed EMP stars were formed as the member of supernova binaries of $M_1 > M_{\rm up}$.  
   For example, for $M_{\rm md}$ in the range obtained in Fig.~\ref{s/nos}, the proportion of supernova binaries accounts for $40 \sim 60 \%$ of the currently observed EMP stars.  
   The proportion of low-mass binaries is less than 0.1 \% and the contribution to EMP stars is less then 1 \%.  
   Note that our conclusion is dependent on the assumption of n(q), which is presently very uncertain. 
   In order to explain the origins of not only CEMP-s stars but also CEMP-nos stars, however, the IMFs of EMP stars have to be weighted in the intermediate and more larger mass range. 
   Otherwise, one has to seek other formation mechanism(s) for all the CEMP-nos stars, which are utterly unknown to the current theory of stellar evolution and/or nucleosynthesis.   

The present results further provide a way of probing into the stellar populations that have left EMP stars now constituting the Galactic halo.  
   We may estimate the total stellar mass necessary to explain the number of currently observed EMP stars.  
   From the above derived IMF, we expect one low-mass star of $M < 0.8 \msun$ out of EMP binaries of number $N_{\rm b} = M_{\rm md}/ 0.8 \msun$, and hence, of total stellar mass $(3/2) N_{\rm b} M_{\rm md}$ on average if we assume the same flat distribution of mass ratio [$n (q) = 1$] as above.   
   On the other hand, from the recent large scale survey, the number of observed EMP stars in the Galactic halo is estimated at $\sim 670 \hbox{ steradian}^{-1}$ for $\feoh < -2.5$ and with the limiting magnitude of $ B \lesssim 17.5$ \citep[][in deriving this number we take the ratio between EMP stars of $ \feoh < -2$ and $ \feoh < -3$ from their Table 3]{Beers05}. 
   With this limiting magnitude, giants can be reached to distance up to $\sim 100$ kpc, and hence, within the whole stellar halo, while dwarfs can be reached only in neighborhood of $\sim 3 $ kpc.  
   Giants occupy about a half among the observed EMP stars, as discussed in \S~4. 
   By taking into account the mass range of stars now on the giant branch, $\Delta M_{\rm G} = 0.01 \msun$ and the flat mass function of EMP stars, we may estimate total number of EMP stars in the Galactic halo, 
\begin{equation}
N_{\rm EMP} \simeq 670 \times 0.5 \times 4 \pi \times 0.8 \msun /\Delta M_{\rm G} \simeq 3.4 \times 10^5. 
\end{equation}
   Thus, the total mass, $M_{\feoh < -2.5}$, of stars in the mother stellar populations of $\feoh < -2.5$ that have produced these low-mass EMP stars that survive to date in the Galactic halo, amounts to be 
\begin{equation}
M_{\feoh < -2.5} \simeq 6 \times 10^7 \msun \left (N_{\rm EMP}/3 \cdot 10^5 \right) \left( M_{\rm md} / 10 \msun \right)^2.
\end{equation}
   The loci of constant mass of $M_{\feoh < -2.5} \ (= 10^7, \ 10^8$, and $10^9 \ \msun)$, obtained numerically from the equations given in the preceding subsections, are plotted in Fig.~\ref{mcsigma} (dash dotted lines).  
   The total mass of stars in the mother populations increases with $M_{\rm md}$, and most of them have exploded as supernovae.  
   Accordingly, the metal production by these erstwhile supernovae may impose an upper bound on the total stellar mass that has been involved in the mother populations, and hence, an upper bound on the medium mass.  
   If we take the averaged iron yield to be $\sim 0.01$ of the initial stellar mass, then, the mother populations may have increased the iron abundance in our Galaxy up to 
\begin{equation} 
\feoh \simeq - 2 + \log \left( M_{\feoh < -2.5} / 10^8 \msun \right)  
\end{equation}
   on an averaged basis all over the Galaxy of total (baryon) mass $M \simeq 10^{11} \msun$. 
   With the derived high-mass IMF, the total mass of stars of $10^8 \msun$ are sufficient to promote the chemical evolution of the Galaxy to the metallicity of population II.  
   As a consequence, the IMFs with $M_{\rm md}$ significantly exceeding $\sim 10 \msun$ may be excluded by the metal overproduction.  

In summary, the EMP stars currently observed originate from a small fraction ($\sim 10\%$ of binary systems in number) of stellar populations of total mass $\sim 10^8 \msun$ that have once constitute, or merged into the Galactic halo.
   A significant portion of them become CEMP stars, and other significant portion ($40 \sim 60$\%) have been exposed to supernova explosion of their companions.  
   In the supernova binaries, the secondary stars are likely to be set unbound after the supernova explosions as a result of sudden reduction of the primary mass. 
   The secondary stars may possibly interact with the envelope matter lost through the wind before being released, and also with SN ejecta from the primary stars.  
  
 We conclude this section with two comments on the implications of these supernova binaries to the observed characteristics of EMP stars.
Because of large wind velocity and the expansion velocity of supernovae, only a small fraction of ejecta can be accreted by the secondary stars, and yet, it may influence the surface characteristics for the elements of such small abundances as the r-process elements.  
   The r-process is most poorly understood among the stellar nucleosynthesis mechanisms, and yet, it may be agued from the solar abundances and the supernova rates that the amount of r-process elements, ejected per one event, is of an order of $M_{\rm r-p} \simeq 10^{-5} \msun$ on average \citep{Mathews90,Woosley94}.  
   Then, simply assuming the geometrical cross sections for the accretion, we may expect the surface enrichment of r-process elements of secondary stars as large as
\begin{eqnarray}
[\abra{r}{Fe}] \simeq  1.3 & + & \log (M_{\rm r-p} /10^{-5} \msun) - 2 \log A ({\rm AU}) \nonumber \\
& -  &\log (M_{\rm scz, \ G}/ 0.35) \msun
\end{eqnarray}
for the giants of $\feoh \simeq - 3$.  
   EMP stars are known to display variations of r-process element abundances with a large range by a factors of $\sim 1000$ of $-1 < [\abra{Eu}{Fe}] < 2$ \citep{Honda04}. 
   The above estimate can be compatible with the observed enrichment, with the largest one from the systems of the smallest separations.  
   In particular, our binary scenario gives a straightforward explanation to the observed large variations in term of the difference in the binary separation.  
   This new channel of surface pollution is worth future investigations with the interactions between the matter ejected by supernova explosion and the secondary star taken into account.  
   It may also happen that the secondary stars accrete the envelope mass ejected by wind from primary stars before the supernova, and also, be polluted through the accretion in gas shell of supernova remnants after the explosion.  

Among the supernova binary systems, the primary stars of $M_{\rm up } < M_1 \lesssim 11 \msun$ have been proposed to make supernova explosion, triggered by electron capture on $^{20}$Ne and oxygen burning in the electron-degenerate O + Ne core by \citet{Miyaji80}. 
   On the other hand, \citet{Ritossa96} show that the helium layer is dredged up by the surface convection during the carbon shell-burning, and then, carbon shell-burning is extinguished and these stars enter into the thermal pulsating (Super) AGB phase with the hydrogen and helium double shell burnings.  
   If this is the case, these stars end in O + Ne white dwarfs, ejecting the envelope through wind mass loss, just as AGB stars with C + O core \citep{Ritossa99,Gil-Pons05}, although lower efficiency of mass loss may tend to narrow the mass range for stars of lower metallicity.  
   Consequently, some of the binaries with the primary stars in this mass range may produce EMP stars of small carbon abundance but largely enriched with nitrogen, similar to ``mixed'' stars, because of dilution in larger envelope mass and of deeper processing by the hot bottom burning on larger core mass at the onset of TP-SAGB phase, as in the upper mass end of white dwarf binary systems of $M < M_{\rm up}$.  
   In addition, because of the very short lifetimes of these massive primaries ($< 10^8$ yr), the secondary stars may suffer from the surface pollution by later accreting interstellar gas, enriched with metals, if their parent clouds persist sufficiently long to be polluted by supernova ejecta of subsequent generations \citep{Suda04}.  

\section{Conclusions and Discussion}\label{conclude}

In this paper, we propose that the origins of carbon-enhanced metal-poor (CEMP) stars, currently observed in the Galactic halo, are explained in terms of the evolution of binary systems of extremely metal-poor (EMP) stars of $\feoh \lesssim -2.5$. 
   Our binary scenario is based on the evolutionary models for both EMP stars and binary mass transfer.  
   We have examined it by confronting the consequences with the observations such as the relation of the carbon-enhancement to the orbital periods and the statistical properties of CEMP stars and also through the comparison with their counterparts of higher metallicity, CH stars. 
    Finally, we demonstrate that a constraint can be imposed on the initial mass functions of EMP stars and discuss the implications to nature of EMP stars in the Galactic halo. 

Main results are summarized as follows: \hfill\break
(I) \ 
   We present an updated summary of evolution of EMP stars. 
   The primary stars can develop the surface carbon enhancement via two distinct mechanisms when evolving to AGB; 
   He-flash driven deep mixing (He-FDDM) in the mass range of $M \lesssim 3.5 \msun$ and third dredge-up (TDU) in the mass range of $ 1.5 \msun \lesssim M \lesssim M_{\rm up}$, where $M_{\rm up}$ denote the upper mass limit of stars ended as white dwarfs.  
   Nitrogen enhancement results either from He-FDDM or from the hot bottom burning (HBB) in the envelope, and s-process elements are synthesized in the helium convective zone promoted by hydrogen mixing precedent to He-FDDM.   
\hfill\break
(II) \ 
   The secondary stars accrete a part of carbon-rich envelope, ejected from the AGB primary stars, to be CEMP stars.
   CEMP-s stars with nitrogen and s-process enhancement originate from the EMP binary systems with the primary stars of mass between $0.8 \lesssim M_1 \lesssim 3.5 \msun$ through He-FDDM.  
   On the other hand, CEMP-nos stars without s-process enhancement stem from the systems with the primary of mass $ 3.5 M_\odot \lesssim M_1 \lesssim M_{\rm up}$. 
   In the latter case, the abundance of nitrogen as well as of carbon vary with the mass of primary, and nitrogen-rich stars with mild or little carbon enhancement are associated with the massive primaries.   
\hfill\break
(III) \ 
   Our binary scenario is shown to give reasonable accounts to the observed characteristics and statistical features of CEMP stars.  
   The large fraction of CEMP stars are explained by the broad parameter space of mass range and binary separation of progenitor binary systems as compared with CH stars.  
   In particular, it is shown that CEMP binary systems have the orbital periods of much wider range than CH stars; 
   all the detected orbital periods and the lack of detected variations in the radial velocity from about a half of CEMP stars are both consistent with the predictions from the binary scenario and from the time span of monitoring of binarity to date.  
\hfill\break
(IV) \ 
   From the observed frequencies of CEMP-s and CEMP-nos stars among EMP stars, we demonstrate that the initial mass function (IMF) of EMP stars has to be massive with the medium mass, $M_{\rm md} \gtrsim 6 \msun$ if approximated to a lognormal form.  
   This also implies that the currently observed EMP stars in the Galactic halo were formed exclusively as the members of binary systems;  
   low-mass EMP stars born as single stars account for a tiny fraction (less than $\sim 1 \%$) if the single stars were born at nearly equal numbers with the similar initial mass function to that of primary stars.  
   Accordingly, in addition to CEMP stars now with white dwarf companions, a significant fraction of EMP stars ($40 \sim 60\%$) used to have a massive primary star and have been exposed to supernova explosion before dismissed from the binary systems. 
   We suggest, as a new channel of surface pollution of EMP stars, the accretion from stellar wind matter and SN ejecta of massive companions.  
\hfill\break
(V) \ 
   From the total number of EMP stars in the Galactic halo, obtained from the recent large-scale surveys, the total mass of stars in their mother stellar populations is estimated to be $\sim 10^8 \msun (M_{\rm md}/10 \msun)^2$.  
   The metal production of SNe imposes another constraint and may rule out the very high mass IMF with the medium mass, $M_{\rm md}$, significantly exceeding $10\msun$.

The following picture of Galactic halo emerges from the present study.  
   Galactic halo has once involved the stellar populations of metallicity $\feoh < -2.5$ and of mean mass $\sim 10 \msun$, which contain the binary systems as many as $\sim 10^7 (M_{\rm md}/10 \msun)$ in total; 
   about a half of them had the primary components of low and intermediate masses that have now evolved to white dwarfs, and another half have more massive primary components that have ended as supernovae. 
   The currently observed EMP stars are $\sim 10 $\% survivors of the low-mass members of these binaries.  
   The mother stellar populations of EMP stars therefore constitute 0.1\% of baryon mass of our Galaxy ($\simeq 10^{11} \msun$). 
   SNe from the massive stars comprised in them suffice to raise the metallicity of whole Galaxy to $\sim 0.01$ solar on average, leading to the formation of population II objects, observed in the Galactic halo.   

Our results indicate that the transition from an IMF dominated by massive stars to an IMF overwhelmed by low-mass stars occurred after the metallicity is raised above $\feoh \sim -2.5$. 
   Our interpretation on the origin of CEMP-nos stars, i.e., TDU and hot bottom burning, should be common to the stars of more metal rich generations.  
   Therefore, the lack of their correspondences in Pop II stars, particularly nitrogen-rich "mixed" stars, may be attributed to the difference in the initial mass function, as discussed in \S \ref{IMFsec}.  
   The metallicity at the transition suggested from the present work seems significantly larger than claimed from the studies on the dynamical and thermal evolution of gas clouds under the metal deficient circumstance.  
   The recent studies tend to argue that the metal and dust cooling can supersede the cooling by hydrogen molecules even for the metallicity as small as $\feoh \sim - 5$, which is argued to reduce the Jeans mass and produce the fragments below a solar mass \citep[e.g., see ][]{Omukai00,Omukai05}.  
   It is also shown that the sub-solar Jeans masses and fragments can be formed even from gas completely devoid of metals, once gas is heated to be ionized by a shock \citep{Uehara00} either due to the collapse of massive primordial objects with total (baryon plus dark matter) mass $M \gtrsim 10^8 \msun$ or due to supernova explosion of the first generation stars \citep{Machida05}.  
   In these studies the Jeans mass is directly connected to the masses of formed stars. 
   It is true however that the process of fragmentation is still poorly understood and a proper understandings of star formation is yet to be established. 
   It is worth noticing here that the globular clusters, which embrace a host of low-mass stars, exist only for the metallicity of $\feoh > -2.5$ in our Galaxy.  

One of the critical assumptions in the present study is on the distribution of mass ratio $n(q)$ of binary systems.  
   It is still an open question either observationally or theoretically, as discussed in \S \ref{population}.   
   We postulate a flat mass function of secondary components extending well below a tenth of the mean mass of primary components, which enables us to explain the origins not only of CEMP-s stars but also of CEPM-nos stars within the current standard framework of the theory of stellar evolution. 
   If both the binary components are assumed to have the same distribution functions, as done by \citet{Lucatello05b}, we should seek for the other origin of CEMP-nos stars presumably elsewhere outside the current theory.  
   In addition, we have pointed out that a flat distribution of mass ratio finds support in the low-mass cut off reported for CH stars \citep{Luck91}. 

Another critical assumption in our argument is on the s-process nucleosynthesis in EMP star.  
   In the present paper, we work out the stellar evolution and nucleosynthesis within the standard frameworks, which takes into account the thermal convection and chemical diffusion as the mechanisms of material mixing in the stellar interior with a negligible contribution of convective overshooting.  
   It is true that there exist the observations of surface abundance anomalies, not explicable within these frameworks.  
   One of the relevant issues is the formation of \nuc{13}C pocket, which are proposed to work as neutron sources in the metal-rich stars, but can not be realized within the current standard framework of stellar evolution. 
   Our scenario postulates that it is inefficient and the s-process nucleosynthesis via radiative \nuc{13}C burning is not effective in the stars of metallicity below $\feoh < -2.5$, at least, in the massive AGB stars of $M \gtrsim 3.5 \msun$.  
   In actuality, the necessity of \nuc{13}C pocket is claimed from the comparisons with the observed distribution of s-process elements in stars of more metal rich populations, and yet, it is treated as a free parameter.  
   The present results may give an insight into the modeling of \nuc{13}C pocket formation if our interpretation of CEMP-nos is correct. 

As for s-process nucleosynthesis in EMP stars, we may point out that the relevance of our high-mass IMFs to the synthesis, in particular, of the light s-process elements.  
   Most of CEMP-nos stars in our samples show larger abundances of strontium than barium ($[{\rm Sr}/{\rm Ba}] \gtrsim 0$), distinct from CEMP-s stars which show larger enrichment of the main and heavy s-process elements relative to the light s-elements.  
   This may be explicable in terms of the difference in the mass of primary stars, and hence, in the core mass when the primary stars start the TP-AGB evolution.  
   For the primary stars of CEMP-nos stars, the core mass can be sufficiently large and the temperature in the helium flash convection high enough to burn \nuc{22}Ne via $\nucm{22}{Ne} (\alpha, n) \nucm{25}{Mg}$, and to promote the s-process nucleosynthesis \citep{Iben75b}.  
   This neutron capture process is likely to produce mostly light s-process elements and to yield such distribution with the ratio of $[{\rm Sr}/{\rm Ba}] > 0$, as shown by \citet{Truran77}, for the very source reaction produces neutron poison, and hence, the available neutron per seed nuclei are restricted to be rather small.  
   Recently \citet{Aoki05} report a general tendency of the excess of light over main s-process elements increasing toward lower metallicity of $\feoh < -2.9$ up to $[\abra{Sr}{Ba}] \simeq 1.5$.  
   The helium shell flashes in massive AGB stars can be one of the candidates for the site of light s-process element synthesis.  

As for some CEMP stars, there have been proposed other origins than the binary evolution; 
   \citet{Umeda03} construct a peculiar supernova model with a carbon-rich ejecta and propose the second generation of stars formed of gas mixed with the ejecta as a possible scenario for CEMP stars.  
   In particular, for stars below $\feoh < -5$, recently discovered, the formation of low mass stars is argued by assuming the high carbon abundance $[\abra{C}{H}] \simeq -1.3$ as observed from the stars \citep{Bromm03}.  
   Since the amount of carbon yield from type II SNe is of $\sim 0.2 \msun $, however, it is open to question whether the second generation stars were formed of carbon abundance as large as $[\abra{C}{H}] = -1.2 \sim -1.3$ with such a small amount of carbon ejected.  
   This seems at an apparent variance to the fact that most of EMP stars have iron abundances of $\feoh \lesssim -2.5$ although they are thought to be formed with iron ejecta of mass $0.1 \sim 1 M_\odot$ in the similar way;  
   Further with the smallest energy of explosion $< 10^{51} \hbox{ erg}$ assigned to the carbon-rich SNe, the star formation triggered by supernova \citep{Tsujimoto99} is unlikely to work since the shock driven by SN dies away before the fragmentation occurs in the swept shell \citep{Machida05}.  
   Even if the stars of the second generation are not formed, however, our binary scenario may propose an alternative channels of the surface pollution of the low-mass members of the first generation by accreting the wind matter and supernova ejecta.  
   This is worth consideration in future works.  

Finally, we comment on the surface pollution by accreting interstellar gas in the parent clouds where these EMP stars were born, discussed by \citet{Shigeyama03} and by \citet{Suda04}. 
   The latter argue that the surface of EMP dwarfs may be polluted up to $\feoh \gtrsim -3$ if the parent clouds survive sufficiently long ($\gtrsim 10^9$ yr) and enrich their gase with metals up to metallicity of $\feoh \simeq -2$ and higher. 
   If the lifetimes of primary stars are shorter than $\lesssim 10^9$ yr as in the cases of white dwarf binaries at the massive end and of supernova binaries, EMP stars may have suffered from the accretion of interstellar gas after the mass transfer, and hence, can disguise their surface abundance with that of interstellar gas.  
   When they evolve to giants, however, the surface abundance is diluted by a factor of $\sim 100$ with the internal matter, accumulated by the mass transfer from the primary stars.  
   Accordingly, some of CEMP-nos stars and ``mixed'' stars, and some from the supernova binaries, may have quite different appearance while they are dwarfs and after they evolve to giant branch.    

The surface pollution by accreting interstellar metal-rich gas has been addressed out in relation to the two most iron-deficient, carbon-enhanced stars of $\feoh <-5$ recent discovered by \citet{Christlieb02} and by \citet{Frebel05}.  
   \citet{Suda04} have suggest the possibility that these stars are Pop III stars which have been polluted by accreting iron-rich interstellar gas, and also, have become CEMP stars through the mass transfer from the erstwhile AGB companion in the close binaries.  
   If they are really Pop III survivors, we may expect that some Pop III stars without carbon enhancement exist, too. 
   These two stars are thought to undergo He-FDDM, and hence, if the Pop III stars were formed under the same IMF as obtained above, then, we may expect $8 \sim 20$ Pop III stars mostly without carbon enhancement ($3 \sim 12$ single stars from the supernova binaries). 
   Since their iron group elements are due to pollution with metal-rich interstellar gas, these stars have to display the surface metallicity of $\feoh \simeq -3 \sim -4$ (with the gravitational settling is taking into account) while they are dwarfs and suffer from the dilution of surface pollution as they evolve to giants and deepen the surface convection.  
   The surface abundances of these pollutants should be similar to those observed from HE0107-5240 and HE1327-2326 for the elements accreted from the interstellar gas since for metallicity of $\feoh \simeq -2$ and higher, the variations in the abundances tend to be small.  
   We may anticipate detection of such stars even current compilations of EMP stars with careful investigations.

\acknowledgments
We are very much benefited from discussion with Dr. Icko, Iben, Jr. and Dr. Toshitaka Kajino.  
This work has been partially supported by Grant-in-Aid for Scientific Research (15204010, 16540213, 18104003), from the Japanese Society for the Promotion of Science. 

\appendix

\section{An attempt to Interpret the abundance anomalies from EMP stars}

\subsection{CEMP-nos stars with Oxygen enhancement}

The hot bottom burning may burn not only carbon but also oxygen into nitrogen in metal-poor stars since the temperature in the bottom of surface convection increases with decreasing the metallicity \citep[e.g.,][]{Ventura01}. 
   If this is the case, the abundance of oxygen now observed should be inherent in the secondary stars, mixed and diluted with matter transferred from the primary stars. 
   The carbon abundances larger than the equilibrium ratio to nitrogen may also be attributed to the pristine matter of secondary stars. 
   All CEMP-nos stars listed in our sample are giants and the accreted matter has been mixed in the deep convection.   
   As a corollary, before developing the deep convection, these CEMP-nos stars should have a different appearance as dwarfs that only nitrogen is enhanced; 
   in addition, the surface abundance of dwarfs may also be subject to the pollution by accreting interstellar matter after the mass transfer because of relatively short lifetime of primary stars, as discussed by \citet{Suda04}.   
   It is to be noted that the nitrogen enrichment in CEMP-nos stars is different from that observed from some giants in the globular clusters, attendant with the depletion of oxygen, and hence, the deep mixing mechanisms along the giant branch as proposed for the latter \citep[e.g., see][]{Suda06} will not be applicable. 

\subsection{Nitrogen-rich EMP stars and the Relation to CEMP-nos Stars}

Some EMP stars show large nitrogen enhancement but a little or no carbon enhancement. 
   A well-known example is CD$-38^\circ245$, which were thought as the most metal-poor star until the discovery of HE0107-5240 \citep[$\feoh = -4.5$;][but revised later to $\feoh = -3.98$ by Norris et al.~2002]{Bessel84}, with the abundances of $[\abra{C}{Fe}] = 0-0.3$, $[\abra{N}{Fe}] = 1.7$, $[\abra{O}{Fe}] = 1.3$, and  $[\abra{Sr}{Fe}] \simeq [\abra{Ba}{Fe}] = -0.5$ \citep[][cf. Spite et al.2005]{Bessel87}.  
   For comparison, we plot this star in Fig.~\ref{BaFeNCFe} with two more similar stars, CS22878-101 ($\feoh = -3.25$) and CS22952-015 ($\feoh = -3.43$).  
   These stars may arise from the same mechanism as CEMP-nos stars with nitrogen enrichment but may belong to the binaries with more massive primary and of wider separation, yielding greater processing by hot bottom burning and smaller enhancement of $[\abra{C+N}{Fe}]$.  
\citet{Spite05} observe EMP giants with weak or no carbon enhancement and argue that they are grouped into ``mixed'' stars with $[\abra{N}{C}] \simeq 1$, including above three stars, and ``unmixed'' stars of $[\abra{N}{Fe}] < 0.5$, both with similar $[\abra{C+N}{Fe}]$.  
   They compile 17 ``mixed'' stars, among which 4 stars have $[\abra{C+ N}{Fe}] > 0.5$, and our list of CEMP-nos stars shares one star, CS22949-037, of $[\abra{C}{Fe}] > 0.5$.
   For the ``mixed'' stars, the problem is also reduced to identify the site(s) for conversion of carbon into nitrogen, and they suggest several possible sites, including the hot bottom burning in massive AGB stars and the hydrogen burning in the envelope of very massive stars. 
   In the latter case, however, we have to seek a way to retain such large abundance ratios of nitrogen to carbon and to oxygen as observed from the ``mixed'' stars if these stars were formed from matter polluted with matter ejected by supernova explosions of the massive stars since the nitrogen-rich matter from their envelope suffer from dilution due to the mixing with carbon- and/or oxygen-rich ejecta from the inner parts as well as with the interstellar matter \citep[but see][]{Norris02}. 
   Instead, the origin of ``mixed'' stars may well be understood in terms of the binary evolution of case IV$^\prime$ but with the primary of large masses and of large separations.  

\subsection{Abundance Features of CS22892-052 and Relation with Orbital Period}

\begin{figure}
\epsscale{1.0} \plotone{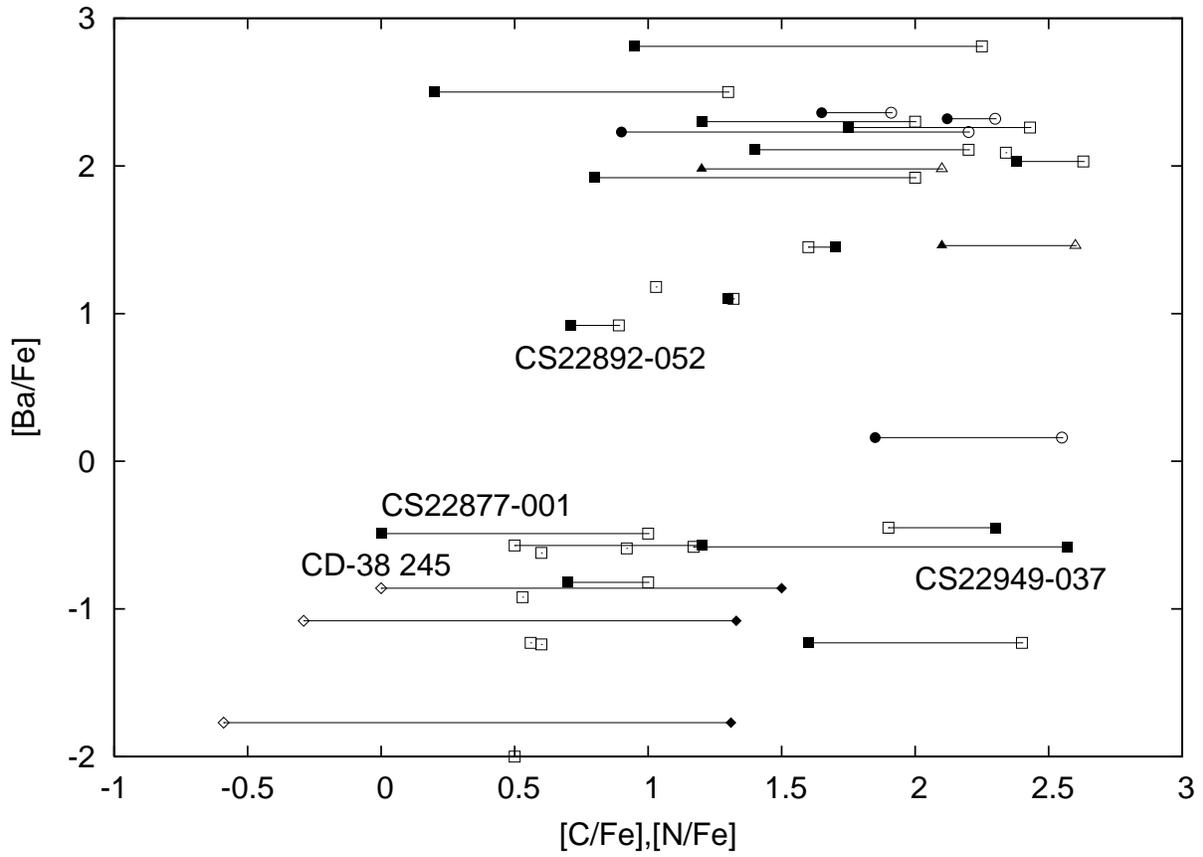}
\caption{Enhancement of s-process elements $[\abra{Ba}{Fe}]$, plotted against the enrichment of carbon (open symbols) and of nitrogen (filled symbols) for CEMP stars. 
   Squares, circles, and triangles denote giants, subgiants, and dwarfs, respectively; diamonds plot three ``mixed'' stars enriched with nitrogen but not with carbon for comparison.  
}\label{BaFeNCFe}
\end{figure}

\begin{figure}
\epsscale{0.65} \plotone{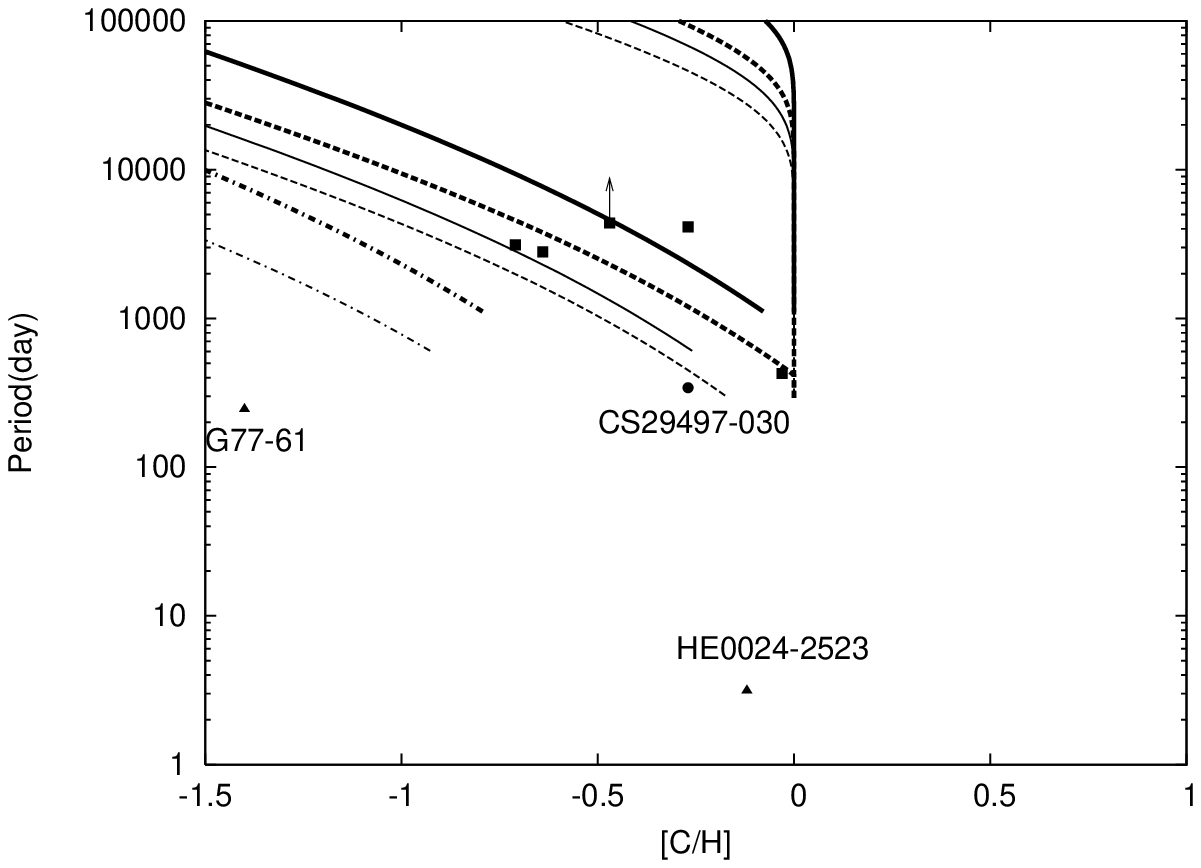} \plotone{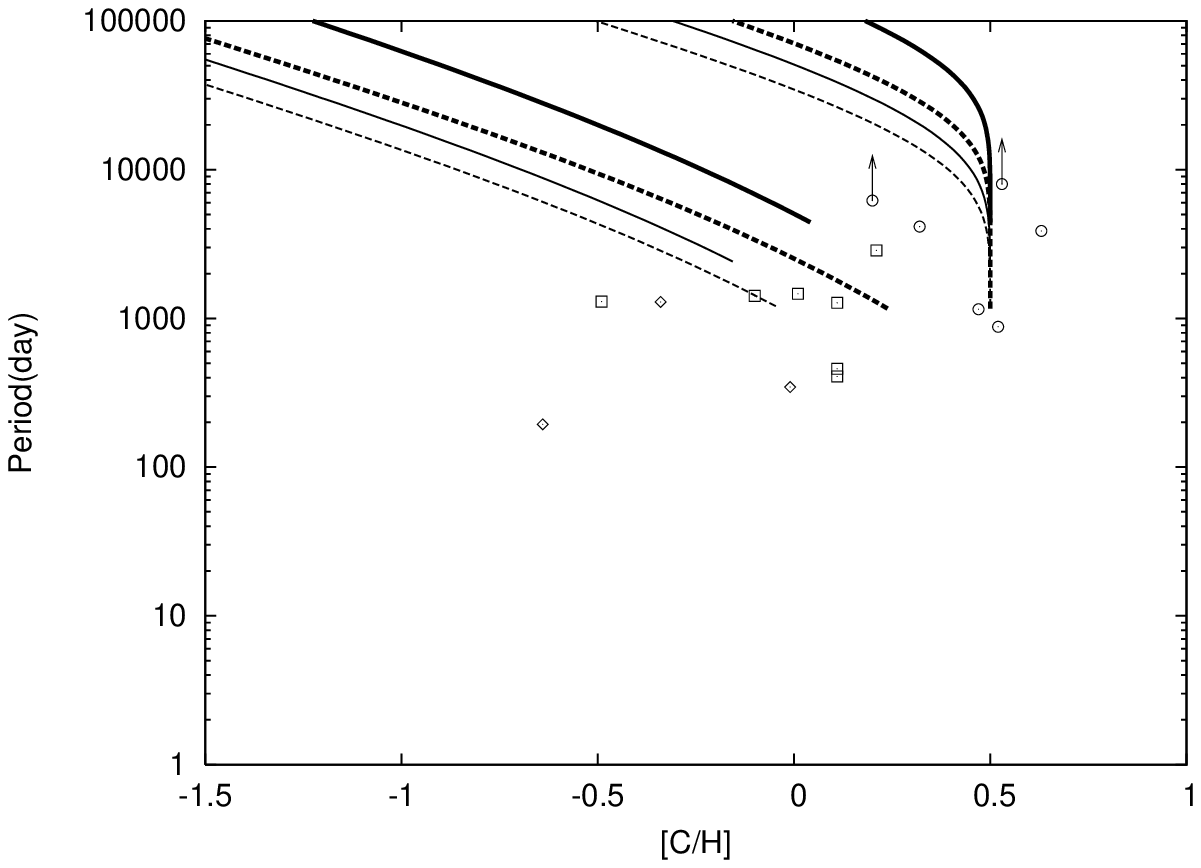}
\caption{Orbital periods vs.\ surface carbon abundance [C/H] for CEMP stars (top panel) and for CH stars (bottom panel).  The theoretical predictions from the wind accretion model described in \S\ref{wa} are plotted under the assumption of Jeans' theorem (solid lines) and of constant separations (broken lines) for giants (on the left-hand side) and for dwarfs and sub-giants (on the right-hand side) with the primary masses of $1.5 \msun$ (thin lines) and $3.0 \msun$ (thick lines).  The mass of the secondary star is set at 0.8 $M_\odot$ and the mass in the surface convection is taken to be $0.0035 \msun$ and $0.35 \msun$ for CEMP dwarf and giants and $0.01 \msun$ and $0.35 \msun$ for subgiant and giant CH stars, respectively. 
   The carbon abundance in the wind is assumed to be $[{\rm C}/ {\rm H}]_{\rm p} = 0$ for CEMP stars and $[{\rm C}/ {\rm H}]_{\rm p} = 0.5$ for CH stars.  
   Two dash-dotted lines in upper panel plot the relations for a wholly convective secondary star of mass $0.3 \msun $.
  Symbols denote observational data for CEMP giants (filled squares), subgiants (filled circle) and dwarfs (filled triangles) in upper panel and for CH stars (open squares), subgiant CH stars (open circles) and field blue stragglers (open diamonds) in lower panel. 
}
\label{cemp-orbit}\label{ch-orbit}
\end{figure}

CS22892-052 is of metallicity $[\abra{C}{H}] \simeq -2$ and observed to show the orbital period of $P = 127$ days. 
   In order to explain the origin in terms of binary scenario, this star seems to require the small carbon abundance $[{\rm C}/{\rm H}]_{\rm p} \lesssim -1$ in the wind, as seen from the comparison with the plots in Fig.~\ref{cemp-orbit}.  
   The binarity of this system is subject to suspicion, however, since it displays radial velocity variations of only a very low amplitude \citep[$\sim 1 \hbox{ km s}^{-1}$;][]{Preston01}. 
   If this star really belong to a binary of $P = 127$ days, the low amplitude implies either a significantly small inclination angle of the orbital plane and/or a very small mass ratio.  
   In the latter case, the companion star is likely to be a brown dwarf rather than a white dwarf, and we have to seek the origin of carbon and other heavy elements of this star in the other site such as the preceding supernovae. 
   More specifically, \citet{Tsujimoto01} have argued that this star with enhanced abundances of r-process elements inherited its heavy elements from the supernova explosion of a $\sim 20\, M_\odot$ star.  On the other hand, this star shows the enhancement and anomalous relative abundances of CNO abundances, as seen from Fig.~\ref{BaFeNCFe}, and this makes the mass transfer in a binary more likely as long as CN enrichment is concerned.


\clearpage
\begin{deluxetable}{lrrrrrrrrrrl}
\tabletypesize{\normalsize}
\rotate
\tablecaption{Observational Sample of CEMP Star from Literature}
\tablewidth{0pt}
\tablehead{\colhead{ Object } & \colhead{ Period } & \colhead{ e } & \colhead{ log g } & 
\colhead{[Fe/H]} & \colhead{[C/Fe]} & \colhead{[N/Fe]} & \colhead{[O/Fe]} & 
\colhead{[Sr/Fe]} & \colhead{[Ba/Fe]} & \colhead{[Pb/Fe]} & \colhead{ Ref.} 
}
\startdata
\sidehead{stars with radial velocity variation}
HE0024-2523 & 3.14 & 0 & 4.3 & -2.72 & 2.6 & 2.1 & 0.40 & 0.34 & 1.46 & 3.3 & 1 \\
G77-61 & 245 & 0 & 5.05 & -4.03 & 2.6 & 2.6 & / & 0 & $<$1 & / & 2 \\
CS29497-030 & 342 & 0 & 4.1 & -2.57 & 2.30 & 2.12 & 1.48 & 0.84 & 2.17 & 3.55 & 3,4,5 \\
CS22948-027 & 426.5 & 0.02 & 1.8 & -2.47 & 2.43 & 1.75 & / & 0.90 & 2.26 & 2.72 & 6,7,8 \\
CS22942-019 & 2800 & 0.1 & 2.4 & -2.64 & 2.0 & 0.8 & / & 1.7 & 1.92 & $<$1.6 & 8,9,10 \\
CS22957-027 & 3125 & 0.45 & 2.4 & -3.11 & 2.4 & 1.6 & / & -0.56 & -1.23 & / & 8,9,10 \\
CS29497-034 & 4130 & 0.02 & 1.8 & -2.90 & 2.63 & 2.38 & / & 1.00 & 2.03 & 2.95 & 7 \\
LP625-44 & $>$12yr & / & 2.5 & -2.72 & 2.25 & 0.95 & 1.85 & 1.32 & 2.81 & 2.55 & 11,12,13 \\
CS22892-052 & 127.8? & ? & 1.6 & -3.03 & 0.89 & 0.71 & 0.72 & 0.44 & 0.92 & 1.2 & 8,13,14,15,16 \\
CS30301-015 & binary \tablenotemark{b}  & / & 0.8 & -2.64 & 1.6 & 1.7 & / & 0.3 & 1.45 & 1.7 & 9,10 \\
CS29526-110 & binary & / & 3.2 & -2.38 & 2.2 & 1.4 & / & 0.88 & 2.11 & 3.3 & 9,10 \\
HE2148-1247 & binary & / & 3.9 & -2.3 & 1.91 & 1.65 & / & 0.76 & 2.36 & 3.12 & 17 \\
CS22877-001 & binary? & / & 2.2 & -2.85 & 1.0 & 0.0 & / & -0.12 & -0.49 & / & 6,18,19 \\
CS22183-015 & binary? & / & 2.5 & -2.85 & 2.34 & / & / &  & 2.09 & 3.17 & 19,20 \\
\sidehead{stars with s-process enhancement}
HE1327-2326 & / & / & ? & -5.4 & 4.1 & 4.5 & $<$4.0 & 1.1 & $<$1.4 & / & 21 \\
HE0107-5240 & / & / & 2.2 & -5.28 & 3.70 & 2.57 & 2.3 & -0.52 & $<$0.82 & / & 22,23 \\
HE0007-1832 & / & / & 3.8 & -2.65 & 2.55 & 1.85 & / & 0.14 & 0.16 & $<$2.82 & 24 \\
CS22947-187 & / & / & 1.3 & -2.49 & 1.03 & / & / & 0.57 & 1.18 & / & 25 \\
CS31062-012 \tablenotemark{a}  & / & / & 4.5 & -2.55 & 2.1 & 1.2 & / & 0.3 & 1.98 & 2.4 & 9,10 \\
HD187216 & / & / & 0.4 & -2.48 & 1.3 & 0.2 & / &  & 2.5 & / & 26 \\
CS31062-050 & / & / & 3 & -2.31 & 2 & 1.2 & / & 0.91 & 2.30 & 2.9 & 9,10,27 \\
CS22898-027 & / & / & 3.7 & -2.26 & 2.2 & 0.9 & / & 0.92 & 2.23 & 2.84 & 9,10 \\
HD196944 & / & / & 1.8 & -2.25 & 1.32 & 1.3 & / & 0.84 & 1.10 & 1.9 & 9,10 \\
\sidehead{stars with no s-process enhancement}
CS22949-037 & / & / & 1.5 & -3.9 & 1.17 & 2.57 & 1.98 & 0.33 & -0.58 & / & 15,28,29 \\
CS29498-043 & / & / & 0.6 & -3.75 & 1.9 & 2.3 & 2.9 & -0.35 & -0.45 & / & 10 \\
CS22891-200 & / & / & 1 & -3.5 & 0.53 & / & / & -1.33 & -0.92 & / & 23 \\
CS22897-008 & / & / & 1.5 & -3.41 & 0.56 & / & / & 0.63 & -1.23 & / & 15,24 \\
BS16929-005 & / & / & 2.7 & -3.09 & 0.92 & / & / & 0.28 & -0.59 & / & 14 \\
CS30325-094 & / & / & 2.1 & -3 & 0.5 & / & / & -2.35 & $<$-1.76 & / & 30 \\
CS29516-041 & / & / & 2.5 & -3 & 0.5 & / & / & -1.95 & -2.00 & / & 29,30 \\
CS30314-067 & / & / & 0.7 & -2.85 & 0.5 & 1.2 & / & -0.37 & -0.57 & / & 6 \\
BS16033-008 & / & / & 2.7 & -2.8 & 0.6 & / & / & -0.68 & -1.24 & / & 30 \\
CS30325-028 & / & / & 1.8 & -2.8 & 0.6 & / & / & 0.27 & -0.62 & / & 30 \\
CS29502-092 & / & / & 2.1 & -2.76 & 1 & 0.7 & / & -0.4 & -0.82 & / & 6 \\
\enddata
\tablecomments{We select CEMP star as [Fe/H] $\lesssim$ -2.5 and [C/Fe] $\geq $ 0.5 from literature.  
Also included are N rich star with [Fe/H] $\leq $ -2.2 and [C/Fe] $\geq $ 0.5. }
\tabletypesize{\scriptsize}
\footnotesize{
\tablenotetext{a}{CS31062-012=LP706-7. }
\tablenotetext{b}{Radial velocity variations are observed, but not enough to estimate period.}
\tablerefs{
(1) \citet{Lucatello03}; 
(2) \citet{Plez05}; 
(3) \citet{Sivarani04}; 
(4) \citet{Ivans05};
(5) \citet{Sneden03};
(6) \citet{Aoki02a}; 
(7) \citet{Barbuy05}; 
(8) \citet{Preston01};  
(9) \citet{Aoki02b}; 
(10) \citet{Aoki02c}; 
(11) \citet{Aoki01}; 
(12) \citet{Aoki02d}
(13) \citet{Norris97b}; 
(14) \citet{Honda04};
(15) \citet{Spite05};
(16) \citet{Sneden03};
(17) \citet{Cohen03};
(18) \citet{Giridhar01}; 
(19) \citet{Tsangarides04}; 
(20) \citet{Johnson02};
(21) \citet{Frebel05}; 
(22) \citet{Christlieb02}; 
(23) \citet{Christlieb04}; 
(24) \citet{Cohen04}
(25) \citet{McWilliam95}; 
(26) \citet{Kipper94}; 
(27) \citet{Johnson04};
(28) \citet{Depagne02}; 
(29) \citet{Norris01}; 
(30) \citet{Aoki05b}
}}\label{table}
\end{deluxetable}

\label{tab1}

\begin{deluxetable}{lrrrrrrrl} 
\tabletypesize{\scriptsize  }
\tablecaption{Observational Sample of CH or Ba Star from Literature}
\tablewidth{0pt}
\tablehead{\colhead{ Object } & \colhead{  binarity } & \colhead{ Period (days) } &
\colhead{  [Fe/H] } & \colhead{  [C/H] } & \colhead{ [C/Fe] } & \colhead{ [Ba/Fe] } & \colhead{class} & \colhead{ Ref.}}
\startdata
HD77247 & y & 80.55 & / & / & / & / & Ba & 1 \\
HD46407 & y & 458.6 & -0.42 & 0.11 & 0.53 & 1.39 & Ba & 1,2 \\
HD199939 & y & 584.9 & / & / & / & / & Ba & 1 \\
HD58368 & y & 672.7 & / & / & / & / & Ba & 1 \\
HD31487 & y & 1066.4 & / & / & / & / & Ba & 1 \\
HD223617 & y & 1301 & / & / & / & / & Ba & 1 \\
NGC2420-X & y & 1402 & -0.7 & / & / & / & Ba & 1,3 \\
HD13611 & y & 1642.1 & / & / & / & / & Ba & 4,5 \\
HD101013 & y & 1710.9 & / & / & / & / & Ba & 1 \\
HD49641 & y & 1768 & / & / & / & / & Ba & 1 \\
HD16458 & y & 2018 & / & / & / & / & Ba & 1 \\
HD204075 & y & 2422 & / & / & / & / & Ba & 1 \\
HD205011 & y & 2853 & / & / & / & / & Ba & 1 \\
HD178717 & y & 2866 & 0 & 0.21 & 0.21 & 0.7 & Ba & 1,3,6 \\
HD131670 & y & 2948 & / & / & / & / & Ba & 1 \\
HD196673 & y & 4000 & / & / & / & / & Ba & 1 \\
HD199394 & y & 4390 & / & / & / & / & Ba & 1 \\
HD139195 & y & 5324 & / & / & / & / & Ba & 5,7 \\
HD202190 & y & 6489 & / & / & / & / & Ba & 5,7 \\
HD5825 & / & / & -0.7 & -0.85 & -0.15 &  & Ba & 8 \\
HD4084 & / & / & -0.7 & -0.58 & 0.12 &  & Ba & 8 \\
HD15589 & / & / & -0.7 & -0.45 & 0.25 &  & Ba & 8 \\
HD27271 & / & / & -0.5 & -0.35 & 0.15 &  & Ba & 8 \\
HD44896 & / & / & -0.4 & 0.36 & 0.76 & 0.82 & Ba & 3,6 \\
HD104979 & n & / & -0.33 & -0.38 & -0.05 & / & Ba & 1,9 \\
HD83548 & / & / & -0.33 & 0.04 & 0.37 & 0.49 & Ba & 10 \\
HD65699 & / & / & -0.3 & -0.45 & -0.15 &  & Ba & 8 \\
HR774 & / & / & -0.3 & 0.31 & 0.61 & 1.03 & Ba & 3,6 \\
HD116713 & / & / & -0.29 & 0.27 & 0.56 & 1.62 & Ba & 2 \\
HD109061 & / & / & -0.21 & -0.32 & -0.11 & 0.6 & Ba & 3 \\
HD95345 & / & / & -0.2 & -0.45 & -0.25 &  & Ba & 8 \\
HD65966 & / & / & -0.2 & -0.17 & 0.03 & 0.49 & Ba & 10 \\
HD60197 & / & / & -0.2 & 0.21 & 0.41 & 0.57 & Ba & 3,6 \\
$\xi$ Cyg & / & / & -0.1 & 0.4 & 0.5 & 0.9 & Ba & 3,11 \\
HD92626 & / & / & 0.04 & 0.68 & 0.64 & / & Ba & 3 \\
HD89638 & / & / & 0.1 & 0.1 & 0 & 0.5 & Ba & 3 \\
$\xi$ Cap & / & / & 0.1 & 0.13 & 0.03 & 1 & Ba & 3,12 \\
HD121447 & / & / & 0.1 & 0.31 & 0.21 & 0.57 & Ba & 3,6 \\
HD100012 & / & / & 0.33 & -0.41 & -0.74 & 0.28 & Ba & 3 \\
$+42^\circ$  2173 & y & 328.3 & / & / & / & / & CH & 1 \\
HD209621 & y & 407.4 & -0.9 & 0.11 & 1.01 & / & CH & 1,13 \\
$+08^\circ$  2654A & y & 571.1 & / & / & / & / & CH & 1 \\
HD5223 & y & 755.2 & / & / & / & / & CH & 1 \\
HD224959 & y & 1273 & -1.6 & 0.11 & 1.71 & / & CH & 1,13 \\
HD198269 & y & 1295 & -1.4 & -0.49 & 0.91 & 2.07 & CH & 1,13,14 \\
HD135148 & y & 1416 & -1.88 & -0.1 & 1.78 & / & CH & 1,15 \\
HD201626 & y & 1465 & -1.3 & 0.01 & 1.31 & / & CH & 1,13 \\
HD30443 & y & 2954 & / & / & / & / & CH & 1 \\
HD187861 & / & / & -1.65 & 0.41 & 2.06 & / & CH & 13 \\
$-38^\circ$  2151 & / & / & -1.4 & -0.39 & 1.01 & / & CH & 13 \\
HD189711 & / & / & -1.15 & 0.80 & 1.95 & 1.80 & CH &16 \\
HD42272 & / & / & -1.10 & 1.21 & 2.31 & 1.65 & CH & 16 \\
HD197604 & / & / & -0.90 & 0.85 & 1.75 & 1.80 & CH & 16 \\
HD25408 & / & / & -0.82 & 0.70 & 1.52 & 1.47 & CH & 16 \\
HD59643 & / & / & -0.70 & 0.58 & 1.28 & 2.95 & CH & 16 \\
HD26 & n & / & -0.44 & 0.01 & 0.45 & / & CH & 1,13,17 \\
HD104340 & / & / & -1.15 & -0.62 & 0.53 & -0.83 & MDBa\tablenotemark{b} & 3 \\
HD204613 & y & 878 & -0.35 & 0.52 & 0.87 & 0.56 & SGCH & 17,18,19 \\
HD89948 & y? & 1153 & -0.27 & 0.47 & 0.74 & 0.83 & SGCH & 17,18,19 \\
HD122202 & y & 1290 & -0.1 & / & / & / & SGCH & 18 \\
BD $+17^\circ$  2537 & y & 1796 & -0.2 & / & / & / & SGCH & 18 \\
HD202020 & y & 2046 & -0.1 & / & / & / & SGCH & 18 \\
HD216219 & y & 3871 & -0.32 & 0.63 & 0.95 & 0.89 & SGCH & 17,18,19 \\
HD11377 & y & 4140 & -0.05 & 0.32 & 0.37 & 0.03 & SGCH & 17,18,19 \\
HD4395 & y? & $>$6200 & -0.33 & 0.2 & 0.53 & 0.56 & SGCH & 17,18,19 \\
HD182274 & y? & $>$8000 & -0.18 & 0.53 & 0.71 & 0.59 & SGCH & 17,18,19 \\
HE0143-0441 & / & / & -2.16 & -0.5 & 1.66 & 2.36 & SGCH & 20 \\
CS22881-036 & n & / & -2.06 & -0.10 & 1.96 & 1.93 & SGCH & 21 \\
CS22880-074 & n & / & -1.76 & 0.25 & 1.51 & 1.34 & SGCH & 21 \\
HD107574 & / & / & -0.8 & -0.03 & 0.77 & 0.74 & SGCH & 22 \\
HD87080 & / & / & -0.51 & -0.14 & 0.61 & 1.51 & SGCH & 23 \\
HD123585 & / & / & -0.5 & 0.37 & 0.87 & 0.82 & SGCH & 3 \\
HD76225 & / & / & -0.5 & 0.18 & 0.68 & 0.94 & SGCH & 22 \\
HD141804 & / & / & -0.41 & -0.17 & 0.24 & 0.84 & SGCH & 3 \\
HD88446 & / & / & -0.36 & 0.02 & 0.38 & 0.64 & SGCH & 13,17,18,19 \\
HD219116 & / & / & -0.34 & 0.34 & 0.68 & 0.9 & SGCH & 19 \\
HD50264 & / & / & -0.34 & 0.41 & 0.59 & 1.25 & SGCH & 23 \\
HD92545 & / & / & -0.33 & -0.07 & 0.26 & 0.52 & SGCH & 22 \\
HD176021 & / & / & -0.3 & 0.34 & 0.64 & / & SGCH & 17 \\
HD188985 & / & / & -0.3 & 0.06 & 0.36 & 0.95 & SGCH & 22 \\
HD150862 & / & / & -0.3 & 0.12 & 0.42 & 0.67 & SGCH & 22 \\
HD125097 & / & / & -0.16 & 0.66 & 0.82 & 0.75 & SGCH & 14 \\
CS29509-027 & y & 194 & -2.02 & -0.64 & 1.38 & 1.33 & FBS\tablenotemark{a} & 24 \\
CS29497-030 & y & 346 & -2.16 & -0.01 & 2.15 & 2.45 & FBS & 24 \\
CS22956-028 & y & 1290 & -2.08 & -0.34 & 1.74 & 0.37 & FBS & 24 \\
\enddata
\tablecomments{}
\tablenotetext{a}{FBS: Field Blue Straggler }
\tablenotetext{b}{MDBa: Metal-deficient barium star }
\label{CHtable}
\tablerefs{
(1) \citet{McClure90}; 
(2) \citet{Kovacs85}; 
(3) \citet{Luck91}; 
(4) \citet{Griff85};
(5) \citet{Boffin94}
(6) \citet{Smith84}; 
(7) \citet{Griff91}; 
(8) \citet{Barbuy92}; 
(9) \citet{Luck91b}; 
(10) \citet{Kovacs83}; 
(11) \citet{Chromey69};
(12) \citet{Smith80}; 
(13) \citet{Venture92}; 
(14) \citet{Lee74};
(15) \citet{Carney03}; 
(16) \citet{Kipper96};
(17) \citet{Luck82}; 
(18) \citet{McClure97}; 
(19) \citet{Smith93}; 
(20) \citet{Cohen04}; 
(21) \citet{Preston01};
(22) \citet{North94}; 
(23) \citet{Pereira03};
(24) \citet{Sneden03}  
}
\end{deluxetable}

\label{tab2}
\clearpage

\clearpage

\clearpage

\clearpage

\clearpage

\clearpage

\clearpage

\clearpage

\clearpage

\clearpage

\clearpage

\clearpage

\end{document}